\documentclass[journal, twoside]{IEEEtran}
\usepackage[table]{xcolor}
\usepackage{indentfirst}
\usepackage{graphicx}
\usepackage{amsmath}
\usepackage{amssymb}
\usepackage{amsfonts}
\usepackage{mathrsfs}
\usepackage{leftidx}
\usepackage{color}
\usepackage{amsmath}
\usepackage{arydshln}
\usepackage{amsthm}
\usepackage{ragged2e}
\usepackage{cite}
\usepackage{enumerate}
\usepackage{longtable}
\usepackage{float}
\usepackage{stfloats}
\usepackage{hyperref }
\usepackage{algpseudocode}
\usepackage{algorithm}
\usepackage[caption=false,font=normalsize]{subfig}
\usepackage{tabularx}
\usepackage{booktabs} 
\usepackage{multirow}

\theoremstyle{plain}

\newcommand{\RMnum}[1]{\uppercase\expandafter{\romannumeral #1}}

\usepackage{caption}

\newcolumntype{P}[1]{>{\raggedright\arraybackslash\footnotesize}m{#1}}
\newcolumntype{A}[1]{>{\centering\arraybackslash\footnotesize}m{#1}}

\usepackage[table,usenames,dvipsnames]{xcolor}
\definecolor{aa}{RGB}{175,238,238}
\definecolor{bb}{RGB}{255,255,255}

\usepackage{bm}
\usepackage{makecell}

\begin{document}

\title{SemDPLA: Semantic Communication-based Distributed Physical-Layer Authentication for 6G-enabled Dense IoT}

\author{Rui Meng,~\IEEEmembership{Member,~IEEE,} Xiqi Cheng, Song Gao, Yuankang Chen, Yinqiu Liu,~\IEEEmembership{Member,~IEEE,} 

Xiaodong Xu,~\IEEEmembership{Senior Member,~IEEE,} Pei Xiao,~\IEEEmembership{Senior Member,~IEEE,} 
Rahim Tafazolli,~\IEEEmembership{Fellow,~IEEE,}

and Ping Zhang,~\IEEEmembership{Fellow,~IEEE} 

\thanks{
This work was supported in part by the National Key Research and Development Program of China under Grant 2020YFB1806905; in part by the National Natural Science Foundation of China under Grant 62501066 and under Grant U24B20131; in part by the S\&T Program of Hebei under Grant 262X0405D; and in part by the Long Term Science and Technology Plan for Broadcasting, Television, and Online Audiovisual Program under Grant 2025AD0300.

Rui Meng and Xiaodong Xu are with State Key Laboratory of Networking and Switching Technology, Beijing University of Posts and Telecommunications, Beijing 100876, China, and also with the Satellite Internet Testing Center, Xiong'an Aerospace Information Research Institute, Xiong'an 070001, China (e-mail: buptmengrui@bupt.edu.cn; xuxiaodong@bupt.edu.cn).

Xiqi Cheng, Song Gao, Yuankang Chen and Ping Zhang are with the State Key Laboratory of Networking and Switching Technology, Beijing University of Posts and Telecommunications, Beijing 100876, China (email: chengzi@bupt.edu.cn; wkd251292@bupt.edu.cn; ykchen@bupt.edu.cn; pzhang@bupt.edu.cn).

Yinqiu Liu is with the College of Computing and Data Science, Nanyang Technological University, Singapore 639798 (email: yinqiu001@e.ntu.edu.sg).

Pei Xiao and Rahim Tafazolli are with Institute for Communication Systems (ICS), Home for 5GIC \& 6GIC, University of Surrey, Guildford, Surrey, GU2 7XH, United Kingdom (email: p.xiao@surrey.ac.uk; r.tafazolli@surrey.ac.uk).

}}

\maketitle

\begin{abstract}
With the rapid development of 6G, increasingly dense device connectivity imposes more strict requirements on multi-users Physical-Layer Authentication (PLA). Compared with cryptography-based methods, PLA enables lightweight authentication by using the uniqueness of wireless channels. However, existing PLA schemes in dense wireless scenarios often suffer from weak fingerprint discriminability and limited computation and communication resources. To address these challenges, we propose a Semantic Communication-based Distributed PLA (SemDPLA) framework. The framework constructs fused central semantic Channel State Information (CSI) fingerprints by fusing semantic information from a central node and multiple distributed nodes. Specifically, we introduce semantic communication to reduce the impact of low Signal Noise Ratio (SNR) and the consumption of communication resource during the transmission from distributed nodes to central node.  Furthermore, we propose an ArcFace-based classification method and a semantic fingerprint-oriented distributed voting consistency mechanism to enhance device classification accuracy. Simulation results demonstrate that the proposed SemDPLA scheme performs better than single-node authentication, decision fusion, raw-CSI transmission, and feature fusion baselines. It achieves equal error rates (EERs) of 8.6\% at 0 dB and 3.4\% at 20 dB. It also achieves classification accuracies of 91.4\% at 0 dB and above 95.8\% from 5 to 20 dB. Moreover, SemDPLA is robust in low SNR environment and under attacks of abnormal nodes.

\end{abstract}


\section{Introduction}


\subsection{Background}


As a key usage scenario of 6G, massive communication has drawn wide attention from both academia and industry \cite{9369324}. It aims to push the boundaries of connection density, with a target density of $10^7$ devices per square kilometre, roughly an order of magnitude higher than the 5G massive machine-type communication (mMTC) standard \cite{saad2019vision}. This trend is driving traditional Internet of Things (IoT) toward dense IoT, which will comprise of vast numbers of low-cost, low-power, and resource-constrained devices. Therefore, dense IoT will impose greater demands on network access, identity authentication, and security management \cite{ghaleb2026lightweight}\cite{scalise2024systematic}. 


In the current 5G-enabled IoT, device authentication primarily relies on encryption mechanisms, such as 5G AKA and EAP AKA' implemented in the core network \cite{r5}. However, applying these directly to 6G-enabled dense IoT presents several challenges, including the security limitations posed by attackers' enhanced computational power, the difficulty for low-resource IoT devices to manage increased complexity, and compatibility issues in heterogeneous and decentralized networks \cite{cheng2026apeg}.

Serving as a robust complement to cryptography-based authentication schemes, Physical-Layer Authentication (PLA) realizes lightweight and low latency identity authentication by leveraging unique characteristics derived from communication links and location-specific attributes at the physical layer \cite{lu2020reinforcement}, such as Channel State Information (CSI) \cite{zhang2025physical,meng2023physical} and Radio Frequency (RF) fingerprints \cite{r9}. These inherent features serve as distinctive and dynamic fingerprints, providing natural protection and personalized identification for devices.
In particular, CSI fingerprints rely on the principle of spatial decorrelation in the channel, enabling them to effectively distinguish devices separated by distances greater than half a wavelength\cite{r11}. Furthermore, CSI fingerprints are highly sensitive to environmental conditions and exhibit significant time-varying characteristics, thereby enhancing the system’s resistance to replay attacks \cite{r12}. 


Currently, many researchers have developed centralized PLA (CPLA) schemes, such as statistical decision methods based on the Neyman–Pearson lemma to construct optimal decision boundaries \cite{r16, ji2025physical}, machine learning methods based on SVM \cite{r20} and GMM \cite{r22} to learn the CSI fingerprint classifier, and deep learning methods to extract high-dimensional fingerprint features \cite{r24, 10428002}.

However, traditional CPLA operates independently without assistance from other devices, which face several challenges in 6G-enabled dense IoT as follows. \textit{1) Limited identity characteristics:} Due to constraints of a single observation and errors in channel estimation, existing CPLA framework struggles to accurately capture the spatial-temporal-frequency characteristics of transmitters. This limitation impacts the accuracy and robustness of authentication. \textit{2) Single point of failure:} The centralized architecture is susceptible to becoming a target for attackers. The attackers only need to compromise the central node to disrupt the entire authentication process, such as optimal location attacks \cite{forssell2021worst}. \textit{3) Computing and storage pressure:} Centralized processing demands significant computing resources and storage space, especially when supporting complex authentication models. This can potentially create bottlenecks in system design and deployment.

\subsection{Motivations}

To address the above-mentioned limitations of traditional CPLA, Distributed PLA (DPLA) with the assistance of multi-partners has several advantages as follows \cite{forssell2021worst,li2025distributed}. \textit{1) Spatial diversity gain:} By deploying cooperative nodes in different geographic locations, distributed receivers can gather multiple observations of identity attributes. This approach enhances authentication accuracy and resilience against fading. \textit{2) Enhanced robustness:} Attackers face significant challenges when attempting to mimic channel features collected from multiple distributed collaborators in diverse locations. Even if some nodes are compromised or fail, others can continue functioning, thereby bolstering the overall system's resilience and defense against attacks. Several recent studies have investigated distributed PLA from different perspectives. 


Despite the above advantages and recent advances of DPLA, there remains a major challenge when applying existing DPLA schemes directly to 6G-enabled dense IoT: \textbf{how to efficiently transmit the fingerprints obtained by collaborators and fuse them at Bob ?} Firstly, DPLA requires frequent transmission of raw CSI between collaborators and Bob, which consumes significant communication resources \cite{xiao2017phy, zhou2025incentive, forssell2021worst}. Furthermore, as the transmission from different collaborators to Bob is affected by varying degrees of channel fading, the quality of the fingerprints also differs \cite{Zhou2024WeightedVotingCPLA}. Consequently, simple fingerprint fusion with fixed weights fails to fully exploit the information from reliable nodes \cite{10214084}.

\subsection{Contributions}
To cope with these problems, we propose Semantic Communication-based Distributed PLA (SemDPLA), a DPLA scheme based on semantic transmission and fusion. 
Specifically, we utilize semantic communication to significantly reduce the communication resource overhead of cooperating nodes \cite{meng2026semantic,fan2026generative,10319671,zhang2026towards}. Furthermore, by implementing adaptive fusion for the fingerprints received at the central node, we enhance the discriminative power of the fingerprints. In addition, we design a new classifier to support multi-device classification tasks in high-density IoT scenarios. Concurrently, by introducing a collaborative optimization mechanism during the training process, we ensure that the distribution of authentication results for the collaborative node fingerprints, the central node fingerprints, and the fused fingerprints remains consistent, thereby improving overall authentication performance.
Overall, the proposed framework combines distributed CSI acquisition, semantic transmission, adaptive fusion and collaborative training. Rather than simply increasing the number of receiving nodes, the framework learns to selectively fuse reliable distributed semantic fingerprints. Consequently, the framework is capable of enhancing the stability and robustness of physical-layer authentication under conditions such as dense connectivity, inconsistent node reliability and semantic link degradation.

The main contributions are summarized as follows.

\begin{enumerate}
  \item We propose a SemDPLA framework for 6G-enabled dense IoT. By transforming multi-node CSI observations into compact semantic fingerprints, the proposed framework reduces raw CSI transmission overhead and supports unified authentication at Bob.
  \item We further design a spatial-frequency semantic encoder for task-oriented CSI fingerprint extraction. The encoder captures antenna-domain, subcarrier-domain, and local spatial-frequency coupling patterns, while attention refinement is used to preserve identity-sensitive CSI semantics.
  \item To effectively aggregate distributed semantic observations, we propose an adaptive semantic fusion mechanism at Bob. By assigning sample-dependent confidence weights to different semantic inputs, the proposed module constructs a unified semantic fingerprint under heterogeneous node reliability and semantic-link conditions.
  \item Moreover, we develop an angular-margin-based authentication and classification module. It improves the separability of legitimate device identities in the fused embedding space and supports an authentication-first and classification-second inference procedure.
  \item We introduce a fusion-centered collaborative training strategy to jointly optimize the above modules. By jointly optimizing fused classification, center-guided semantic alignment, and fusion-guided distribution distillation, the proposed strategy enhances discriminability, cross-node consistency, and cooperative decision making.
\end{enumerate}




\section{System Overview}
In this section, we provide a systematic model of the scene elements, CSI fingerprint acquisition, semantic transmission, fusion fingerprint construction, and authentication modules.

\subsection{Scenario Model}
\begin{figure}[t]
  \centering
  \includegraphics[width=0.5\textwidth,trim=0 0 0 0,clip]{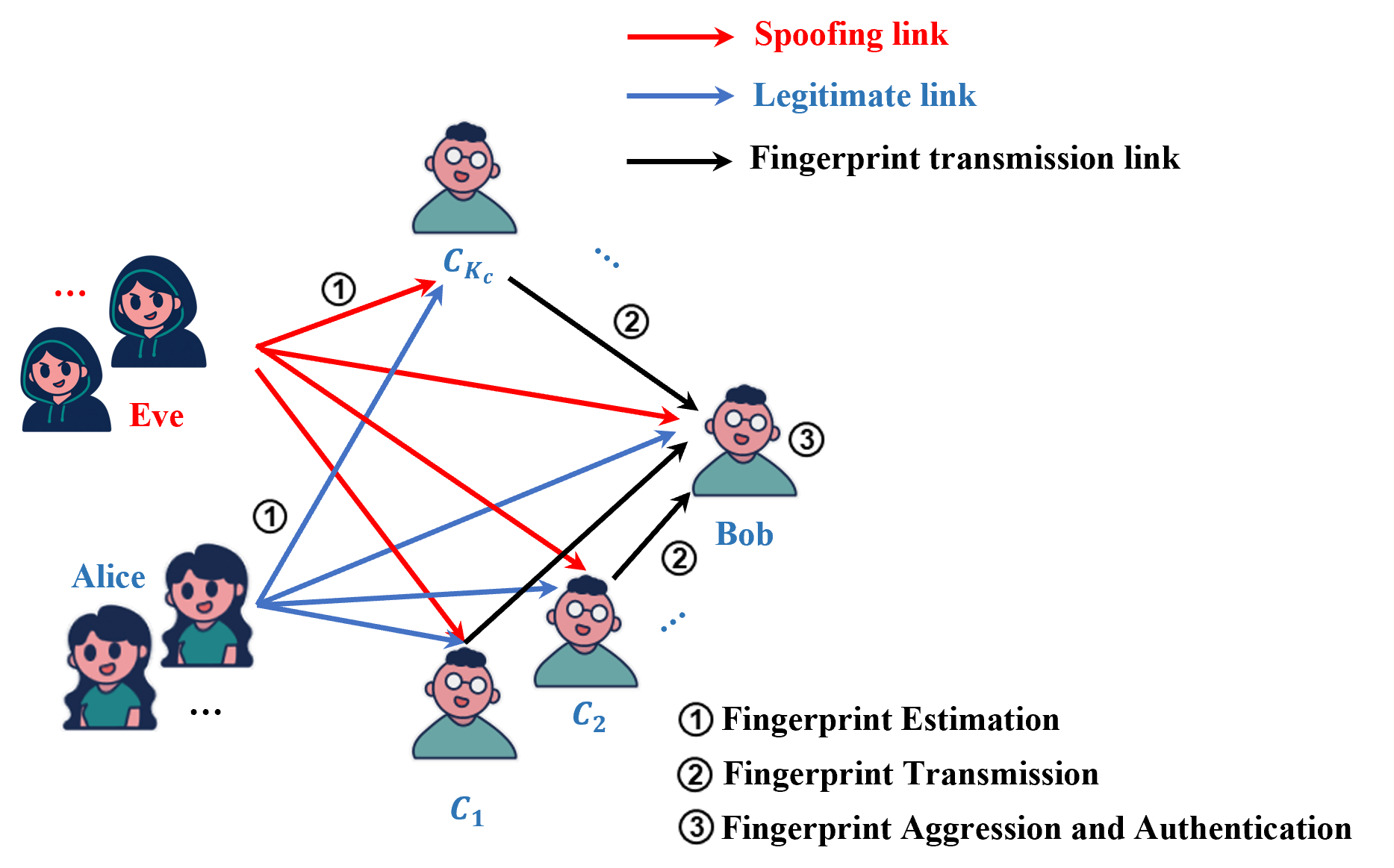}
  \caption{System model of the proposed collaborative authentication architecture, where Bob is the central receiver, Eve is a spoofing attacker, and $Q$ collaborators provide additional observation links for reliable authentication.}
  \label{Authenication system}
\end{figure}

As shown in Fig. \ref{Authenication system}, we consider a dense IoT scenario, where the following nodes are involved:

\begin{itemize}
  \item \textit{Alices ($a_1, a_2, \ldots, a_{S}$):}\hspace{0.3em}Legitimate IoT devices that request access through identity authentication.
  \item \textit{Bob ($c_0$):}\hspace{0.3em}The legitimate center receiver that authenticates multiple IoT devices.
  \item \textit{Collaborators ($c_1, c_2, \ldots, c_{Q}$):}\hspace{0.3em}Distributed nodes that receive signals and transmit extracted fingerprint features to Bob.
  \item \textit{Eves ($e_1, e_2, \ldots, e_{P}$):}\hspace{0.3em}Spoofing attackers that attempt to pretend legitimate devices or disrupt the delivery of CSI fingerprints.
\end{itemize}

In the high-density IoT scenario, legitimate devices are densely deployed. We denote the set of legitimate devices by $\mathcal{A}=\{a_1,a_2,\ldots,a_S\}$, where the devices are randomly distributed within the coverage of Bob and its collaborator. Bob serves as a central receiver $ c_ 0 $ and interacts with $ Q $ collaborators. The set of collaborators is $\mathcal {} = \{c_ 1, c_ 2, \ldots, c_ Q\}$.  Each collaborator receives uplink signals from unknown devices, extracts fingerprint features from local CSI, and forwards them to Bob for feature fusion, device authentication, and multi-device classification. Let $\mathcal{D}$ denote the set of distances from the legitimate devices to Bob and the collaborators, and let $d$ denote the maximum distance between any two legitimate devices. We suppose that $D_{\max} \gg d$, where $D_{\max}=\max \mathcal{D}$. The set of spoofing attackers is denoted by $\mathcal{E}=\{e_1,e_2,\ldots,e_P\}$. Eve acts within Bob's coverage area and can either impersonate legitimate devices by transmitting CSI fingerprints or attack collaborators to degrade the reliability of semantic features during transmission.



\subsection{CSI Acquisition Model}
To characterize the channel from IoT devices to the receivers, we consider that each IoT device is equipped with \(M\) antennas arranged as a uniform linear array (ULA), while Bob is equipped with an \(N\)-element ULA. Let \(h_{m,n}(f,\tau)\) denote the channel impulse response from the \(m\)th transmitter antenna to the \(n\)th receiver antenna at carrier frequency \(f\). The resulting multipath channel is given by \cite{liu2023exploiting}

\begin{equation}
  \label{h_{m,n}}
  h_{m,n}(f,\tau) = \sum\limits_{l=1}^L \alpha_l e^{-j2\pi f\tau_l + j\varphi_l}\delta(\tau-\tau_l),
\end{equation}

\noindent where $L$ denotes the number of propagation paths, $\alpha_l$ represents the gain of the $l$th path, $\varphi_l$ denotes its phase and $\tau_l$ is the corresponding frequency-independent propagation delay. For a wideband orthogonal frequency division multiplexing system, the channel coefficient on the $k$th subcarrier can be expressed as

\begin{equation}
  \label{H_{m,n}}
  H_{m,n}(f, k) = \sum\limits_{l=1}^L \alpha_l e^{-j2\pi f\tau_l + j\varphi_l}e^{-j2\pi \tau_l k/K},
\end{equation}

\noindent where $f_{k}$ denotes the frequency of the $k$th subcarrier relative to the center frequency $f$, and $K$ is the total number of subcarriers. For each receiver antenna $n$, the channel response across the transmit-antenna and subcarrier dimensions can be organized into an $M\times K$ matrix $\bm{H}_{n}(f)$, whose $m$th row is $[H_{m,n}(f,0),\ldots,H_{m,n}(f,K-1)]$. The CSI fingerprint received at collaborator $c_q$ from device $a_s$ is denoted by $\bm{H}_{sq}$. Accordingly, the received signal at node $c_q$ is expressed as
\begin{equation}
  \label{eq:y_asq}
  \bm{y}_{a_s}= \bm{h}_{sq}\bm{x}_s+ \bm{n}_{sq},
\end{equation}

\noindent where $\bm{x}_s$ denotes the transmitted signal matrix of device $a_s$, $\bm{h}_{sq}$ denotes the instantaneous time-frequency domain channel response matrix from device $a_s$ to node $c_q$, and $\bm{n}_{sq}\sim\mathcal{CN}(\bm{0},\sigma_{sq}^2\mathbf{I})$ denotes the complex additive white Gaussian noise vector.

In practice, $c_q$ does not directly obtain the perfect CSI $\bm{h}_{sq}$ and instead relies on an estimate. Therefore, the CSI fingerprint used in this paper is the estimated channel response matrix $\hat{\bm{H}}_{sq}\in\mathbb{C}^{M\times K}$, where the $m$th row $\hat{\bm{h}}_{sq,m}$ contains the estimated instantaneous frequency-domain CSI over the $K$ subcarriers associated with the $m$th transmit antenna. Channel estimation is performed from pilot observations using the least-squares criterion because of its low computational complexity and efficient implementation. During the pilot-transmission stage, the received pilot at node $c_q$ from device $a_s$ is modeled as
\begin{equation}
  \label{pilot_obs_model}
  \bm{Y}_{sq}=\bm{H}_{sq}\bm{X}_{s}+\bm{N}_{sq},
\end{equation}

\noindent where $\bm{X}_{s}$ denotes the known pilot matrix, $\bm{Y}_{sq}$ denotes the corresponding received pilot matrix, and $\bm{N}_{sq}$ is the pilot-domain noise matrix. The LS estimate is obtained by minimizing the cost function, i.e.,
\begin{equation}
  \label{LS_cost_function}
  \mathcal{J}(\bm{H}_{sq})=\left\|\bm{Y}_{sq}-\bm{H}_{sq}\bm{X}_{s}\right\|_{F}^{2}.
\end{equation}
Then, we can obtain the estimated CSI by 
\begin{equation}
  \label{LS_channel_estimator}
  \hat{\bm{H}}_{sq}^{\mathrm{LS}}=\arg\min_{\bm{H}_{sq}}\mathcal{J}(\bm{H}_{sq})=\bm{Y}_{sq}(\bm{X}_{s})^{H}\left(\bm{X}_{s}(\bm{X}_{s})^{H}\right)^{-1}.
\end{equation}
The estimated CSI fingerprint used for authentication is then extracted from $\hat{\bm{H}}_{sq}^{\mathrm{LS}}$. In the single-pilot-vector case, this estimate can be expressed compactly as

\begin{equation}
  \label{eq:h_as_est}
  \hat{\bm{h}}_{sq} = \left( \bm{x}_s^H \bm{x}_s \right)^{-1} \bm{x}_s^H \bm{y}_{a_s},
\end{equation}

\begin{figure*}[tbp]
  \centering
  \includegraphics[width=0.9\textwidth]{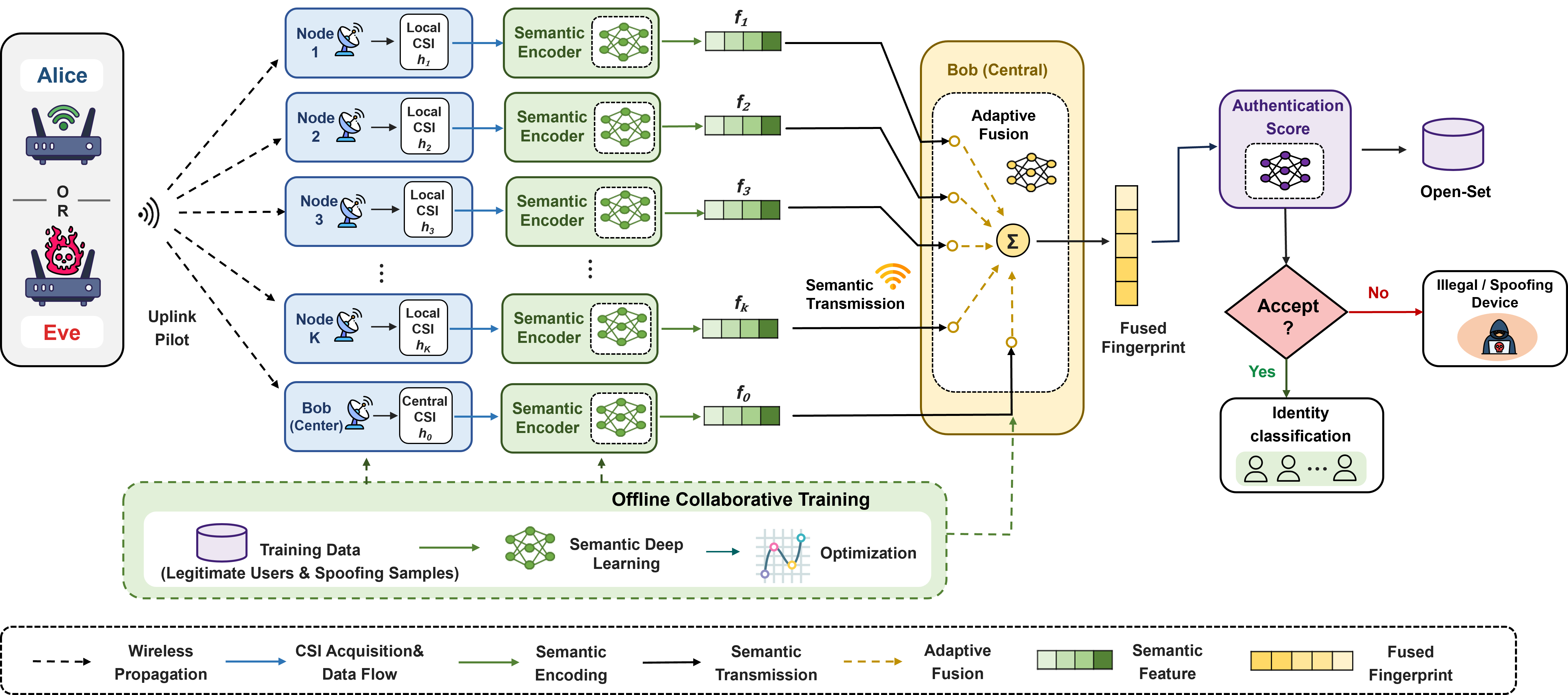}
  \caption{Illustration of the proposed collaborative authentication architecture, where Bob is the central receiver, Eve is a spoofing attacker, and the $Q$ collaborators provide additional observation links for reliable authentication.}
  \label{Authenication system_specifically}
\end{figure*}


\noindent where $\bm{x}_s^H$ denotes the Hermitian transpose of $\bm{x}_s$, and $\hat{\bm{h}}_{sq}$ denotes the estimated CSI fingerprint of device $a_s$ at node $c_q$. Similarly, the signal received at Bob or any collaborator from Eve $e_p$ is expressed as
\begin{equation}
  \label{y_e}
  \bm{y}_{e_p}= \bm{h}_{pq}\bm{x}_p+ \bm{n}_{pq},
\end{equation}

\noindent where $\bm{x}_p$ denotes the transmitted signal matrix (or vector) of Eve $e_p$, $\bm{h}_{pq}$, for $q\in\{0,1,\ldots,Q\}$, denotes the instantaneous time--frequency-domain channel response matrix from Eve $e_p$ to receiver node $c_q$, and $\bm{n}_{pq}\sim\mathcal{CN}(\bm{0},\sigma_{pq}^2\mathbf{I})$ denotes the corresponding complex AWGN vector. The corresponding LS-based estimated CSI fingerprint of Eve can be expressed as

\begin{equation}
  \label{eq:h_ep_est}
  \hat{\bm{h}}_{pq} = \left( \bm{x}_p^H \bm{x}_p \right)^{-1} \bm{x}_p^H \bm{y}_{e_p},
\end{equation}

\noindent where $\bm{x}_p^H$ denotes the Hermitian transpose of $\bm{x}_p$, and $\hat{\bm{h}}_{pq}$ denotes the estimated CSI fingerprint of Eve $e_p$ at node $c_q$.

\subsection{Semantic Transmission and Fusion Model}

To reduce the overhead of raw CSI exchange, we model the communication from collaborators to Bob as semantic transmission. As illustrated in Fig.~\ref{Authenication system_specifically}, we assume that a pre-trained semantic model is deployed at each collaborator $c_q$ and Bob $c_0$. Each collaborator transmits its CSI fingerprint through semantic communication. For collaborator $c_q$, the local semantic CSI fingerprint is given by

\begin{equation}
  \label{eq:local_semantic_representation}
  \bm{z}_{sq}=f_{\theta_q}(\hat{\bm{h}}_{sq}), \quad q\in\{1,\ldots,Q\},
\end{equation}

\noindent where $\bm{z}_{sq}\in\mathbb{R}^{d_z}$ denotes the low-dimensional semantic CSI feature extracted by node $c_q$ from its local estimated CSI fingerprint, and $d_z\ll\dim(\hat{\bm{h}}_{sq})$. Unlike traditional semantic communication schemes that aim to fully reconstruct the raw CSI, $\bm{z}_{sq}$ preserves only the discriminative semantic information required for collaborative fusion and authentication, rather than all channel information.

Before transmission from node $c_q$, the extracted feature is normalized to satisfy the transmit-power constraint of the semantic link. The normalized feature is given by

\begin{equation}
  \label{eq:semantic_power_norm}
  \bar{\bm{z}}_{sq}=\frac{\bm{z}_{sq}}{\sqrt{\frac{1}{d_z}\|\bm{z}_{sq}\|_2^2+\epsilon}},
\end{equation}

\noindent where $\epsilon$ is a small positive constant for numerical stability. The normalized semantic feature is then forwarded to Bob $c_0$ from node $c_q$. The received semantic signal is modeled as

\begin{equation}
  \label{eq:semantic_channel}
  \bm{y}_{sq}^{\mathrm{sem}} = g_q \bar{\bm{z}}_{sq} + \bm{n}_{q}^{\mathrm{sem}},
\end{equation}

\noindent where $\bm{y}_{sq}^{\mathrm{sem}}$ denotes the semantic signal received by Bob from the $q$th collaborator, $g_q$ denotes the channel coefficient of the semantic communication link from $c_q$ to $c_0$, and $\bm{n}_{q}^{\mathrm{sem}}\sim\mathcal{N}(\bm{0},\sigma_{q,\mathrm{sem}}^2\mathbf{I})$ denotes the corresponding AWGN vector. In this way, the collaborator does not transmit the original CSI fingerprint directly, but transmits semantic CSI fingerprints.

After receiving the semantic signals from all collaborators, Bob performs semantic fusion and authentication based on $\{\bm{y}_{sq}^{\mathrm{sem}}\}_{q=1}^{Q}$. Therefore, collaboration among different nodes is carried out in the semantic domain, which reduces transmission overhead while improving the quality of channel features.

\subsection{Central Semantic Fusion and Authentication}

Based on \eqref{eq:semantic_channel}, Bob receives $Q$ semantic CSI fingerprints forwarded by collaborators. Meanwhile, Bob directly extracts a local semantic CSI feature from its own estimated CSI fingerprint $\hat{\bm{h}}_{s0}$, which is expressed as $\bm{z}_{s0}=f_{\theta_0}(\hat{\bm{h}}_{s0})$. $\bar{\bm{z}}_{s0}$ is normalized following ~\eqref{eq:semantic_power_norm}. We define the semantic CSI fingerprint set collected at Bob as $\mathcal{V}_{s}=\left\{\bar{\bm{z}}_{s0},\bm{y}_{s1}^{\mathrm{sem}},\ldots,\bm{y}_{sQ}^{\mathrm{sem}}\right\}$.


Bob then performs semantic aggregation over $\mathcal{V}_{s}$ to obtain a fused semantic CSI fingerprint for authentication and classification. Let $\mathcal{F}_{\phi}(\cdot)$ denote the semantic fusion mapping. The fused CSI fingerprint is denoted as $\bm{u}_{s}=\mathcal{F}_{\phi}(\mathcal{V}_{s})$. The fused representation is further mapped into a discriminative embedding space through a feature projection mapping $\bm{e}_{s}=\mathcal{P}_{\eta}(\bm{u}_{s})$. Based on the projected embedding, Bob generates the decision score vector through the authentication head $\mathcal{C}_{\omega}(\cdot)$ as $\hat{\bm{o}}_{s}=\mathcal{C}_{\omega}(\bm{e}_{s})$. Therefore, the posterior distribution over the legitimate device classes conditioned on the fused semantic CSI fingerprint $\bm{u}_{s}$ can be expressed as 




\begin{equation}
  \label{eq:central_auth_posterior}
  p(\ell\mid \bm{u}_{s})=\frac{\exp\left([\hat{\bm{o}}_{s}]_{\ell}\right)}{\sum_{j=1}^{S}\exp\left([\hat{\bm{o}}_{s}]_{j}\right)}, \quad \ell\in\{1,\ldots,S\},
\end{equation}

\noindent where $[\hat{\bm{o}}_{s}]_{\ell}$ is the score for the $\ell$th legitimate device. Bob selects the device class with the highest probability as the predicted identity $\hat{s}$ and obtain the authentication result as

\begin{equation}
  \label{eq:central_auth_classification}
  \hat{s}=\arg\max_{\ell\in\{1,\ldots,S\}} p(\ell\mid \bm{u}_{s}).
\end{equation}

The softmax output characterizes only the relative posterior support among the legitimate classes and therefore serves as a closed-set classification probability. To further distinguish legitimate devices from spoofing attackers in the open-set authentication setting, Bob defines the authentication confidence score as the posterior probability associated with the predicted class, i.e., 

\begin{equation}
  \label{eq:central_auth_confidence}
  \zeta_{s}=p(\hat{s}\mid \bm{u}_{s})=\max_{\ell\in\{1,\ldots,S\}} p(\ell\mid \bm{u}_{s}),
\end{equation}


\noindent where $\zeta_s$ is the confidence score for open-set thresholding. It is computed from the adaptively fused semantic fingerprint, rather than a single noisy observation. This fusion suppresses unreliable or attacked nodes, making $\zeta_s$ more robust for rejecting spoofing attackers. Therefore, the final authentication decision is given by

\begin{equation}
  \label{eq:central_auth_rule}
  \mathbb{I}_{\mathrm{auth}}=
  \begin{cases}
    1, & \zeta_{s}\geq\gamma, \\
    0, & \zeta_{s}<\gamma,
  \end{cases}
\end{equation}

\noindent where $\gamma$ denotes the authentication threshold. $\gamma$ can be calibrated on a validation set by minimizing the equal error rate (EER) or satisfying a target false acceptance/rejection constraint. A smaller $\gamma$ may increase false acceptance of spoofing devices, while a larger $\gamma$ may lead to more false rejections of legitimate devices. When $\mathbb{I}_{\mathrm{auth}}=1$, the device is accepted as a legitimate device and assigned the identity label $a_{\hat{s}}$. Otherwise, it is rejected as a spoofing or unauthorized device. In general, Bob completes semantic fusion, device classification, and authentication in the semantic space without reconstructing the original distributed CSI fingerprints.


\begin{figure*}[tbp]
  \centering
  \includegraphics[width=0.95\textwidth,trim=1.5 0 0 0,clip]{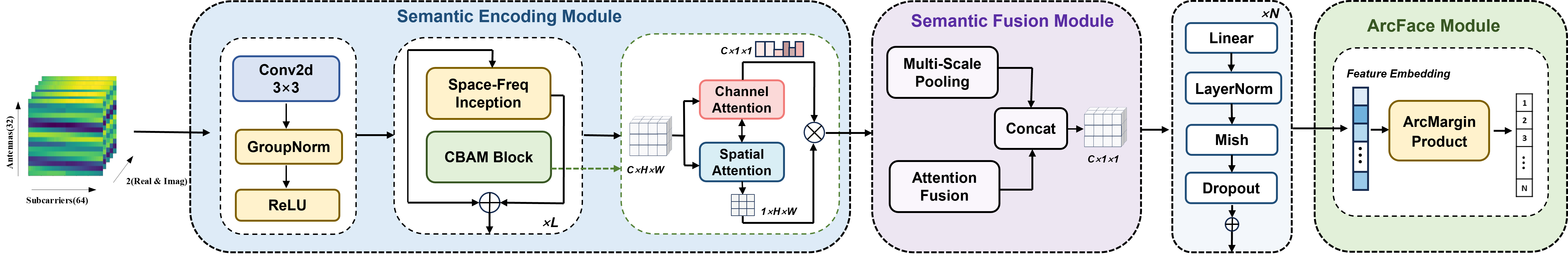}
  \caption{The entire network process, which includes semantic encoding, semantic feature fusion, authentication and classification modules. }
  \label{semdpla_detail}
\end{figure*}

\section{Proposed Distributed Semantic Fusion and Authentication Scheme}

As shown in Fig.~\ref{semdpla_detail}, the proposed SemDPLA framework performs distributed semantic PLA by extracting semantic CSI fingerprints at distributed nodes and adaptively fusing them at Bob. The authentication and classification procedure is implemented through semantic encoding, semantic transmission, adaptive fusion, collaborative training, and online authentication, as summarized in \textbf{Algorithm~\ref{alg:semantic_framework}}.

\subsection{Spatial-Frequency Semantic Encoder}

To extract a discriminative semantic fingerprint from the estimated CSI obtained at distributed receivers, we first rewrite the complex-valued CSI fingerprint $\hat{\bm{H}}_{sq}\in\mathbb{C}^{M\times K}$ as a real-valued two-channel tensor, i.e.,
\setcounter{equation}{16}
\begin{equation}
  \bm{X}_{sq}=[\Re(\hat{\bm{H}}_{sq}),\Im(\hat{\bm{H}}_{sq})]\in\mathbb{R}^{2\times M\times K},
\end{equation}
where the two channels correspond to the real and imaginary parts of CSI, respectively. This representation preserves the full information contained in the complex-valued CSI while allowing the semantic encoder to operate in a standard real-valued convolutional framework.

\begin{algorithm}[t]
  \caption{Overall Procedure of the Proposed SemDPLA Framework}
  \label{alg:semantic_framework}
  \footnotesize
  \begin{algorithmic}[1]
    \State \textbf{Offline encoder pre-training}
    \Repeat{}
    \State Sample a mini-batch of CSI tensors $\{\bm{X}_{sq}\}$
    \State Compute reconstructions $\widetilde{\bm{X}}_{sq}=D_{\varrho}(E_{\theta}(\bm{X}_{sq}))$
    \State Update $(\theta,\varrho)$ by minimizing $\mathcal{L}_{\mathrm{pre}}$ as \eqref{eq:semantic_pretrain_loss_final}
    \Until{$\mathcal{L}_{\mathrm{pre}}$ converges}
    \State Deploy the pre-trained encoder $E_{\theta}(\cdot)$ at Bob and all distributed nodes
    \State \textbf{Offline collaborative training}
    \Repeat{}
    \State Sample a mini-batch of enrolled-device accesses
    \State Extract $\{\bm{v}_{s,q}\}_{q=0}^{Q}$ and obtain $\bm{u}_{s}$ and $\bm{e}_{s}$
    \State Compute $\mathcal{L}_{\mathrm{cls}}$, $\mathcal{L}_{\mathrm{opl}}$, and $\mathcal{L}_{\mathrm{kl}}$ as ~\eqref{eq:training_cls_loss} ~\eqref{eq:training_opl_loss} ~\eqref{eq:training_kl_loss}
    \State Update $(\phi,\eta,\omega,\psi)$ by minimizing $\mathcal{L}$ as ~\eqref{eq:training_joint_loss}
    \Until{$\mathcal{L}$ converges}
    \State \textbf{Online authentication}
    \For{each access attempt}
    \State Form $\{\bm{v}_{s,q}\}_{q=0}^{Q}$ from Bob and distributed observations
    \State Obtain $\bm{u}_{s}$, $\bm{e}_{s}$, and $[\bm{o}_{s}^{\mathrm{inf}}]_c$
    \State Compute $\hat{s}$ and $\zeta_s$, then according to ~\eqref{eq:central_auth_rule}
    \If{$\zeta_s \ge \gamma$}
    \State Accept the device and output identity $a_{\hat{s}}$
    \Else{}
    \State Reject the device as unauthorized
    \EndIf
    \EndFor
  \end{algorithmic}
\end{algorithm}

Since CSI fingerprints exhibit structured correlations over the antenna and subcarrier dimensions, we employ a shared spatial-frequency semantic encoder $E_{\theta}(\cdot)$ in all distributed nodes to perform semantic mapping in a unified way. Its initial feature mapping is written as
\begin{equation}
  \bm{F}_{sq}^{(0)}=\rho\!\left(\mathrm{GN}\!\left(\mathrm{Conv}_{3\times 3}(\bm{X}_{sq})\right)\right),
  \label{eq:semantic_encoder_init_final}
\end{equation}
where $\mathrm{Conv}_{3\times 3}(\cdot)$ denotes a $3\times 3$ convolution, $\rho(\cdot)$ denotes the ReLU activation, and $\mathrm{GN}(\cdot)$ denotes group normalization. Since distributed CSI learning is often carried out with limited batch sizes and node-dependent input statistics, we adopt Group Normalization to reduce the dependence on CSI data statistics and improve the stability of semantic encoding.

\begin{figure*}[!b]
  \hrulefill
  \normalsize
  \setcounter{equation}{20}
  \begin{equation}
    \label{eq:cbam_channel_attention}
    \bm{M}_{c}(\bm{G}_{sq}^{(\ell)})=\sigma\!\left(
      W_{1}\!\left(\rho\!\left(W_{0}(\mathrm{AvgPool}_{sf}(\bm{G}_{sq}^{(\ell)}))\right)\right)
      +W_{1}\!\left(\rho\!\left(W_{0}(\mathrm{MaxPool}_{sf}(\bm{G}_{sq}^{(\ell)}))\right)\right)
    \right),
  \end{equation}
  \setcounter{equation}{22}
  \begin{equation}
    \label{eq:cbam_spatial_attention}
    \bm{M}_{s}(\widetilde{\bm{G}}_{sq}^{(\ell)})=\sigma\!\left(
      f_{7\times 7}\!\left([
          \mathrm{AvgPool}_{c}(\widetilde{\bm{G}}_{sq}^{(\ell)})\,;\,
        \mathrm{MaxPool}_{c}(\widetilde{\bm{G}}_{sq}^{(\ell)})]
    \right)\right).
  \end{equation}
\end{figure*}

To further exploit the relationships of CSI over the antenna and subcarrier dimensions, we introduce a spatial-frequency multi-branch module into the encoder. For the latent feature representation $\bm{F}$, the module is written as
\begin{equation}
  \Psi(\bm{F})=\phi\!\left([\psi_{1\times 1}(\bm{F}),\psi_{1\times 7}(\bm{F}),\psi_{7\times 1}(\bm{F}),\psi_{3\times 3}(\bm{F})]\right),
  \label{eq:spatial_frequency_multibranch_final}
\end{equation}
where $[\cdot]$ denotes channel concatenation, $\psi_{1\times 7}(\cdot)$ captures frequency-domain relationships at the subcarrier dimension, $\psi_{7\times 1}(\cdot)$ captures spatial relationships at the antenna dimension, $\psi_{3\times 3}(\cdot)$ extracts local spatial-frequency features, and $\psi_{1\times 1}(\cdot)$ performs feature compression.

Furthermore, in dense IoT scenarios, the information that distinguishes nearby devices is often concentrated in a small subset of local spatial-frequency regions rather than being uniformly distributed over the whole antenna-subcarrier plane of CSI \cite{kong2023physical}. To emphasize information that is important for authentication, we introduce the Convolutional Block Attention Module (CBAM) \cite{woo2018cbam}. 
Let $\bm{G}_{sq}^{(\ell)}\in\mathbb{R}^{C_{\ell}\times M_{\ell}\times K_{\ell}}$
denote the latent feature tensor at the $\ell$th stage, where $C_{\ell}$, $M_{\ell}$ and $K_{\ell}$ denote the number of channels, the antenna-axis resolution and the subcarrier-axis resolution, respectively.
Specifically, we first apply channel attention to evaluate the importance of different CSI components \cite{woo2018cbam}. By performing global average pooling and global max pooling over the antenna-subcarrier plane and feeding the results into a shared bottleneck mapping, we obtain the channel-attention map as \eqref{eq:cbam_channel_attention},
where $\bm{M}_{c}(\bm{G}_{sq}^{(\ell)})\in\mathbb{R}^{C_{\ell}\times 1\times 1}$ assigns one global weight to each semantic channel, $\sigma(\cdot)$ denotes the sigmoid function, and $W_0$ and $W_1$ denote the shared bottleneck transforms. The channel-refined feature can be expressed as
\setcounter{equation}{21}
\begin{equation}
  \widetilde{\bm{G}}_{sq}^{(\ell)}=\bm{M}_{c}(\bm{G}_{sq}^{(\ell)})\odot\bm{G}_{sq}^{(\ell)}.
  \label{eq:cbam_channel_refined_final}
\end{equation}

Then, we further model the importance of local regions on the spatial-frequency plane. By performing average pooling and max pooling along the channel dimension of $\widetilde{\bm{G}}_{sq}^{(\ell)}$ and feeding the resulting maps into a convolution layer, we obtain the spatial-frequency attention map as \eqref{eq:cbam_spatial_attention},
where $\bm{M}_{s}(\widetilde{\bm{G}}_{sq}^{(\ell)})\in\mathbb{R}^{1\times M_{\ell}\times K_{\ell}}$ assigns one local weight to each antenna-subcarrier position, and $f_{7\times 7}(\cdot)$ denotes a $7\times 7$ convolution. The corresponding spatial--frequency refined feature is denoted as
\setcounter{equation}{23}
\begin{equation}
  \widehat{\bm{G}}_{sq}^{(\ell)}=\bm{M}_{s}(\widetilde{\bm{G}}_{sq}^{(\ell)})\odot\widetilde{\bm{G}}_{sq}^{(\ell)}.
  \label{eq:cbam_spatial_refined_final}
\end{equation}

By enhancing attention features across both the channel and spatial-frequency dimensions, the encoder can compress channel-related features more effectively. In the final semantic transmission stage, the encoder’s output is mapped into a task-oriented semantic representation $\bm{z}_{sq}$, which is then normalized according to \eqref{eq:semantic_power_norm} and transmitted to the central node of Bob.

To provide a more stable initialization for the downstream authentication stage, we pre-train the semantic encoder with a reconstruction-oriented objective before the collaborative training stage. Let $D_{\varrho}(\cdot)$ denote the decoder used only during pre-training. The reconstructed CSI tensor is written as
\begin{equation}
  \widetilde{\bm{X}}_{sq}=D_{\varrho}(E_{\theta}(\bm{X}_{sq})).
  \label{eq:semantic_pretrain_reconstruction_final}
\end{equation}
The pre-training objective is defined as
\begin{equation}
  \mathcal{L}_{\mathrm{pre}}=\frac{1}{N}\sum_{s,q}\left\|\widetilde{\bm{X}}_{sq}-\bm{X}_{sq}\right\|_{F}^{2},
  \label{eq:semantic_pretrain_loss_final}
\end{equation}
where $N$ denotes the number of pre-training samples. After this stage, the pre-trained encoder is transferred to the distributed nodes and the central node as the semantic encoder for feature extraction and semantic transmission.

\subsection{Adaptive Semantic Fusion Scheme}

Based on semantic CSI fingerprint set $\mathcal{V}_{s}$, we consider the central fusion mapping $\mathcal{F}_{\phi}(\cdot)$. The semantic inputs available at Bob are expressed as
\setcounter{equation}{26}
\begin{equation}
  \bm{v}_{s,0}=\bar{\bm{z}}_{s0}, \quad \bm{v}_{s,q}=\bm{y}_{sq}^{\mathrm{sem}}, \quad q\in\{1,\ldots,Q\}.
  \label{eq:adaptive_fusion_inputs_final}
\end{equation}

Since different collaborators may experience different environmental disturbances, their contributions to the final authentication and classification decision should be treated differently. To flexibly fuse distributed semantic CSI fingerprints of varying quality, we design an adaptive semantic fusion mechanism $\mathcal{F}_{\phi}(\cdot)$. Specifically, Bob employs a shared scoring function to evaluate the reliability of each received semantic input for the current authentication sample, which can be expressed as
\begin{equation}
  a_{s,q}=g_{\phi}(\bm{v}_{s,q})
  =
  \bm{w}_{2}^{\top}\tanh(\bm{W}_{1}\bm{v}_{s,q}+\bm{b}_{1})+b_{2}, 
  \label{eq:adaptive_fusion_score_final}
\end{equation}
where $a_{s,q}\in\mathbb{R}$ denotes the relevance score of the $q$th semantic input for the current authentication sample, and $\phi=\{\bm{W}_{1},\bm{b}_{1},\bm{w}_{2},b_{2}\}$ denotes the trainable parameters.

The corresponding fusion weight is obtained by softmax normalization, i.e.,
\begin{equation}
  \alpha_{s,q}=\frac{\exp(a_{s,q})}{\sum_{j=0}^{Q}\exp(a_{s,j})}, 
  \label{eq:adaptive_fusion_weight_final}
\end{equation}
where $\alpha_{s,q}\ge 0$ and $\sum_{q=0}^{Q}\alpha_{s,q}=1$. Hence, $\alpha_{s,q}$ can be interpreted as the relative confidence weight assigned by Bob to the $q$th semantic input.

Accordingly, the fused semantic fingerprint is written as
\begin{equation}
  \bm{u}_{s}=\sum_{q=0}^{Q}\alpha_{s,q}\bm{v}_{s,q},
  \label{eq:adaptive_fusion_output_final}
\end{equation}
which is consistent with the previously defined abstract fusion mapping. Compared with fixed averaging or independent voting, this adaptive fusion mechanism allows Bob to dynamically adjust the weight of each distributed fingerprint according to the quality of the currently received CSI fingerprint, which is therefore more suitable for distributed authentication under different node quality and time-varying channel conditions. Then, the fused CSI fingerprint $\bm{u}_{s}$ is sent to the projection mapping $\mathcal{P}_{\eta}(\cdot)$ to obtain the shared discriminative embedding $\bm{e}_{s}$ for authentication and classification.

\subsection{Authentication and Classification Method}

In dense IoT scenarios, nearby devices may exhibit highly similar CSI fingerprints, making multi-device classification challenging. Semantic communication reduces transmission overhead by transmitting compact CSI fingerprints, while adaptive fusion improves fingerprint reliability by integrating multi-node semantic observations. However, the lightweight fused embeddings may still face class crowding when many legitimate devices are present. To further enhance class separability, we instantiate the central decision mapping $\mathcal{C}_{\omega}(\cdot)$ using an ArcFace-based angular-margin decision head.\footnote{Compared with conventional softmax-based PLA classifiers, ArcFace explicitly enlarges angular inter-class margins, making it more suitable for distinguishing highly similar CSI fingerprints in dense IoT scenarios.} Specifically, ArcFace imposes an additive angular margin between normalized embeddings and class prototypes, thus keeping the lightweight fused semantic fingerprints discriminative without transmitting high-dimensional raw CSI \cite{deng2019arcface}.

For the projected embedding $\bm{e}_{s}$, let $\bm{w}_{c}$ denote the weight vector of the $c$th legitimate device class. The normalized cosine response is expressed as
\begin{equation}
  \cos \theta_{s,c}=\frac{\bm{w}_{c}^{\top}\bm{e}_{s}}{\|\bm{w}_{c}\|_{2}\|\bm{e}_{s}\|_{2}}, \quad c\in\{1,\ldots,S\},
  \label{eq:arcface_cosine_final}
\end{equation}
where $S$ denotes the number of legitimate devices. Let $y_s\in\{1,\ldots,S\}$ denote the identity label of sample $s$ during training. ArcFace introduces an additive angular margin $m$ for the target class as
\begin{equation}
  \phi_{s,y_s}=\cos(\theta_{s,y_s}+m)=\cos\theta_{s,y_s}\cos m-\sin\theta_{s,y_s}\sin m.
  \label{eq:arcface_margin_final}
\end{equation}

By introducing $m$, ArcFace encourages samples within the same class to cluster more closely around their class prototype during training, while widening the angular separation between different classes. Then, the training output of the classification head can be written as
\begin{equation}
  [\bm{o}_{s}^{\mathrm{arc}}]_c=
  \begin{cases}
    \gamma_{a}\,\phi_{s,y_s}, & c=y_s, \\
    \gamma_{a}\,\cos\theta_{s,c}, & c\neq y_s,
  \end{cases}
  \quad c\in\{1,\ldots,S\},
  \label{eq:arcface_train_logits_final}
\end{equation}
where $\gamma_a$ is the scaling factor. Based on these outputs, the closed-set class distribution during training is expressed as
\begin{equation}
  p(c\mid \bm{u}_{s})=\frac{\exp([\bm{o}_{s}^{\mathrm{arc}}]_c)}{\sum_{j=1}^{S}\exp([\bm{o}_{s}^{\mathrm{arc}}]_j)}, \quad c\in\{1,\ldots,S\}.
  \label{eq:arcface_train_posterior_final}
\end{equation}
where $p(c\mid \bm{u}_s)$ is a conditional estimate of the class distribution in the set of legitimate devices. By introducing the angular margin, ArcFace further enlarges the separation among legitimate device classes in the normalized embedding space and alleviates class crowding in dense multi-device scenarios.

During inference, the angular-margin term is removed and the decision head switches to the margin-free form:
\begin{equation}
  [\bm{o}_{s}^{\mathrm{inf}}]_c=\gamma_{a}\cos\theta_{s,c}, \quad c\in\{1,\ldots,S\}.
  \label{eq:arcface_inference_logits_final}
\end{equation}
The device identity is predicted by
\begin{equation}
  \hat{s}=\arg\max_{c\in\{1,\ldots,S\}} [\bm{o}_{s}^{\mathrm{inf}}]_c.
  \label{eq:arcface_identity_prediction_final}
\end{equation}
Then, the open-set authentication score can be expressed as
\begin{equation}
  \zeta_s\triangleq r_s=\max_{c\in\{1,\ldots,S\}} [\bm{o}_{s}^{\mathrm{inf}}]_c=\gamma_a\max_{c\in\{1,\ldots,S\}}\cos\theta_{s,c},
  \label{eq:arcface_authentication_score_final}
\end{equation}
Then, the final authentication decision is expressed as \eqref{eq:central_auth_rule}. The device identity label $\hat{s}$ is output only when $\mathbb{I}_{\mathrm{auth}}=1$.

\section{Proposed Fusion-Centered Collaborative Training Strategy}

To achieve better coordination between fusion and authentication modules, we design a collaborative training strategy for fusion and authentication. We design the training process to achieve three coupled objectives: the fused CSI fingerprint can form clear discriminative boundaries, the distributed semantic fingerprint should remain aligned with the central fingerprint to avoid a bias in the fusion feature center, and the local decisions of distributed fingerprints should stay consistent with the fused global decision. Following this idea, we formulate the overall training objective as

\begin{equation}
  \label{eq:training_joint_loss}
  \mathcal{L}=\mathcal{L}_{\mathrm{cls}}+\lambda_{\mathrm{opl}}\mathcal{L}_{\mathrm{opl}}+\lambda_{\mathrm{kl}}\mathcal{L}_{\mathrm{kl}},
\end{equation}

\noindent where $\mathcal{L}_{\mathrm{cls}}$ establishes the discriminative boundary in the fused semantic space, $\mathcal{L}_{\mathrm{opl}}$ regularizes the geometric alignment between the distributed semantic fingerprints and the central fingerprint. $\mathcal{L}_{\mathrm{kl}}$ further encourages the local decision tendencies of distributed fingerprints to remain consistent with the fused global decision. The training process is summarized in \textbf{Algorithm~\ref{alg:fusion_training}}.

\subsection{Loss Corresponding to Authentication}

To establish the final decision boundary in the unified semantic space, we impose the main classification supervision on the fused branch. Let $y_s$ denote the ground-truth label of the $s$th training sample. Using the ArcFace training logits in \eqref{eq:arcface_train_logits_final}, the main classification loss is defined as

\begin{equation}
  \label{eq:training_cls_loss}
  \mathcal{L}_{\mathrm{cls}}=-\log \frac{\exp\left([\bm{o}_{s}^{\mathrm{arc}}]_{y_s}\right)}{\sum_{j=1}^{S}\exp\left([\bm{o}_{s}^{\mathrm{arc}}]_{j}\right)}.
\end{equation}

\noindent Since the final decision is made from the fused embedding $\bm{e}_s$, we place the main classification supervision on the fused output rather than imposing a separate final classification constraint on each distributed branch, so that the decision boundary is learned directly in the unified semantic space.

\begin{algorithm}[t]
  \caption{Fusion-Centric Collaborative Training Strategy}
  \label{alg:fusion_training}
  \footnotesize
  \begin{algorithmic}[1]
    \Repeat{}
    \State Sample a mini-batch of enrolled-device accesses
    \For{each sample $s$ in the mini-batch}
    \State Obtain $\bm{v}_{s,0}=\bar{\bm{z}}_{s0}$ and $\bm{v}_{s,q}=\bm{y}_{sq}^{\mathrm{sem}}$ for $q=1,\ldots,Q$
    \State Fuse semantics to obtain $\bm{u}_{s}$ and $\bm{e}_{s}$
    \State Compute ArcFace training logits $[\bm{o}_{s}^{\mathrm{arc}}]_c$ and $\mathcal{L}_{\mathrm{cls}}$
    \For{$q=1,\ldots,Q$}
    \State Compute $\mathrm{Proj}_{\bm{v}_{s,0}}(\bm{v}_{s,q})$ and the residual $\bm{v}_{s,q}^{\perp}$
    \EndFor
    \State Evaluate $\mathcal{L}_{\mathrm{opl}}$
    \State Compute $\bm{d}_{s}^{(f)}$ and $\{\bm{d}_{s,q}\}_{q=1}^{Q}$
    \State Evaluate $\mathcal{L}_{\mathrm{kl}}$
    \EndFor
    \State Form $\mathcal{L}=\mathcal{L}_{\mathrm{cls}}+\lambda_{\mathrm{opl}}\mathcal{L}_{\mathrm{opl}}+\lambda_{\mathrm{kl}}\mathcal{L}_{\mathrm{kl}}$
    \State Update $(\phi,\eta,\omega,\psi)$ by back-propagation
    \Until{$\mathcal{L}$ converges}
    \State Return the trained fusion and decision modules
  \end{algorithmic}
\end{algorithm}

\subsection{Loss for Semantic Fusion Alignment}

To regularize the directional deviation of distributed semantic fingerprints after transmission, we use the local semantic fingerprint of central node as a central reference. Let $\bm{v}_{s,0}$ denote the central reference, and $\bm{v}_{s,q}$ denote the distributed semantic fingerprint. Then, the projection of $\bm{v}_{s,q}$ onto $\bm{v}_{s,0}$ can be expressed as

\begin{equation}
  \label{eq:training_opl_projection}
  \mathrm{Proj}_{\bm{v}_{s,0}}(\bm{v}_{s,q})=\frac{\bm{v}_{s,q}^{\top}\bm{v}_{s,0}}{\|\bm{v}_{s,0}\|_2^2}\bm{v}_{s,0}, \quad q\in\{1,\ldots,Q\},
\end{equation}

\noindent where $\mathrm{Proj}_{\bm{v}_{s,0}}(\bm{v}_{s,q})$ denotes the component of $\bm{v}_{s,q}$ parallel to the central reference direction. The corresponding orthogonal residual is then given by

\begin{equation}
  \label{eq:training_opl_residual}
  \bm{v}_{s,q}^{\perp}=\bm{v}_{s,q}-\mathrm{Proj}_{\bm{v}_{s,0}}(\bm{v}_{s,q}),
\end{equation}

\noindent where $\bm{v}_{s,q}^{\perp}$ captures the component deviating from the central reference direction. Based on this decomposition, we write the alignment term as

\begin{equation}
  \label{eq:training_opl_loss}
  \mathcal{L}_{\mathrm{opl}}=\frac{1}{Q}\sum_{q=1}^{Q}\left[1-\frac{\bm{v}_{s,q}^{\top}\bm{v}_{s,0}}{\|\bm{v}_{s,q}\|_2\|\bm{v}_{s,0}\|_2}+\lambda_{\perp}\|\bm{v}_{s,q}^{\perp}\|_2^2\right].
\end{equation}

\noindent In this way, we can keep the distributed semantic fingerprint converging toward the central reference while suppressing orthogonal deviations that are less relevant to the shared semantic content, thereby stabilizing the geometric structure of the cross-node semantic space.

\subsection{Loss for Decision Consistency}

After designing the loss function that aligns feature fusion directions, we introduce a decision alignment mechanism between distributed nodes and the central node during training. By matching the local decisions of the distributed fingerprints with the decisions of the fused fingerprint, we  can improve the accuracy of authentication and classification. We employ a distillation-based distributed learning scheme during training to match the decisions of distributed fingerprints with those of fused fingerprints. Distillation primarily uses the soft output distribution of the teacher model to guide the learning of the student model. Unlike one-hot encoding, the soft distribution not only preserves the correct classification information but also retains the similarity structure between different classes, which can provide richer supervised information for class identification. Specifically, under a temperature coefficient $T$, given an arbitrary output vector $\bm{z}$, the corresponding soft distribution can be expressed as

\begin{equation}
  \label{eq:training_soft_distribution}
  p_i(\bm{z};T)=\frac{\exp\left([\bm{z}]_i/T\right)}{\sum_j\exp\left([\bm{z}]_j/T\right)}.
\end{equation}

Let $\bm{z}^{(t)}$ and $\bm{z}^{(s)}$ denote the teacher and student outputs, respectively. Then, the standard distillation objective is expressed as

\begin{equation}
  \label{eq:training_standard_kd}
  \mathcal{L}_{\mathrm{KD}}=T^2\,\mathrm{KL}\left(p(\bm{z}^{(t)};T)\middle\|p(\bm{z}^{(s)};T)\right),
\end{equation}

\noindent where $\mathrm{KL}(\cdot\|\cdot)$ denotes the Kullback--Leibler divergence. By minimizing the difference between the teacher and student distributions, the student is encouraged to match the class distribution structure of the teacher, following the standard principle of knowledge distillation.

In our framework, given that the fused embedding $\bm{e}_{s}$ contains authentication semantic information from both the distributed nodes and the central node, we use the classification distribution obtained from the fused latent embedding as the teacher signal.
To construct this distribution, we introduce a margin-free auxiliary head $\mathcal{D}_{\psi}(\cdot)$, since the angular margin may change the soft class distribution. 
The auxiliary outputs of the fusion branch and the local node branches are defined as
\begin{equation}
  \label{eq:training_kl_logits}
  \bm{d}_{s}^{(f)}=\mathcal{D}_{\psi}(\bm{e}_{s}), \quad \bm{d}_{s,q}=\mathcal{D}_{\psi}(\mathcal{P}_{\eta}(\bm{v}_{s,q})),
\end{equation}

\noindent
where $\bm{d}_{s}^{(f)}$ denotes the auxiliary output of the fused embedding, $\bm{d}_{s,q}$ is a local class-distribution estimate obtained from the $q$th node feature through the shared projector $\mathcal{P}_{\eta}(\cdot)$ and the auxiliary head $\mathcal{D}_{\psi}(\cdot)$.
Based on these auxiliary outputs, the fusion-guided collaborative distillation loss is given by
\begin{equation}
  \label{eq:training_kl_loss}
  \mathcal{L}_{\mathrm{kl}}=\frac{T^2}{Q}\sum_{q=1}^{Q}\mathrm{KL}\left(\mathrm{Softmax}\left(\frac{\bm{d}_{s}^{(f)}}{T}\right)\middle\|\mathrm{Softmax}\left(\frac{\bm{d}_{s,q}}{T}\right)\right).
\end{equation}

\noindent where $T$ is the temperature parameter. 
$ \mathcal{L}_{\mathrm{kl}}$ guides the distribution estimates to converge towards the distribution based on the fusion fingerprint decision. 
By making the decision embedding distributions of distributed nodes more consistent and reliable, the overall quality of fingerprint fusion is improved and the authentication is more stable.

Overall, the downstream collaborative authentication objective is optimized through fused classification supervision, center-guided alignment, and distribution-level collaborative distillation. The first establishes the discriminative boundary of the fused embedding, the second stabilizes the semantic direction across nodes, and the third further enforces consistency between node-level local distributions and the fused global decision. As a result, the fusion-centered training stage learns a semantic decision space that is jointly discriminative, consistent, and cooperative on top of the pre-trained encoder initialization.

\section{Simulation Results and Analysis}

\subsection{Simulation Settings}

\begin{table}[tbp]\footnotesize
  \caption{Scenario and dataset-construction parameters \cite{Alkhateeb2019}
  \label{tab:table1}}
  \centering
  \begin{tabularx}{\columnwidth}{P{0.47\columnwidth}P{0.43\columnwidth}}
    \toprule
    Parameter & Value\\
    \midrule
    DeepMIMO scenario & Outdoor `O1' \\
    Infrastructure nodes & One central node Bob and $Q=3$ distributed nodes \\
    Device deployment region & Selected rectangular region within the O1 grid \\
    Number of legitimate devices & 100 \\
    Number of attackers & 100 \\
    Time slots per device & 500 consecutive slots \\
    Closed-set identity classes & 100 legitimate devices \\
    Open-set protocol & Legitimate-device authentication against attackers before classification \\
    Training/test split & 4:1 on legitimate-device CSI sequences \\
    Attacker usage & Evaluation only \\
    Temporal phase model & First-order Gaussian--Markov process \\
    Temporal correlation coefficient & $\rho=0.98$ \\
    Phase innovation standard deviation & $3^\circ$ \\
    NLoS Doppler drift & Uniformly distributed within $\pm 10$ Hz \\
    LoS treatment & Dominant LoS path phase-locked when present \\
    CSI preprocessing & First receive antenna retained; antenna/subcarrier dimensions subsampled \\
    Antenna element spacing & $0.5\lambda$ \\
    Radiation pattern & Isotropic \\
    Bandwidth & 50 MHz \\
    Carrier frequency & 28 GHz \\
    Number of propagation paths & 5 \\
    Number of OFDM subcarriers & 32 \\
    \bottomrule
  \end{tabularx}
\end{table}


To emulate a practical dense-IoT deployment, we adopt the outdoor `O1' scenario of the DeepMIMO dataset \cite{Alkhateeb2019} to generate time-varying CSI between the devices in the service area and four receiving nodes, including one central node Bob and $Q=3$ distributed nodes. Specifically, we select a rectangular region from the DeepMIMO grid as the dense deployment area, and then randomly place 100 legitimate devices and 100 attackers in this region while enforcing non-overlapping locations between the two sets. The four nodes jointly provide full communication coverage for the selected area. For each device-node pair, we generate CSI over 500 consecutive time slots to construct temporally correlated channel sequences. Based on the ray-tracing output of DeepMIMO, the path powers and angles are consistent with the dataset, while the path phases evolve across time according to a first-order Gaussian--Markov process with correlation coefficient $\rho=0.98$ and innovation standard deviation $3^\circ$. In addition, a small per-path Doppler drift uniformly distributed within $\pm 10$ Hz is injected into the non-line-of-sight components to emulate slow environmental dynamics, while the dominant LoS path, if present, remains phase locked. After channel construction, only the first receive antenna is retained and the frequency-domain CSI is subsampled along the antenna and subcarrier dimensions to form CSI fingerprints for semantic processing. The key scenario and dataset-construction parameters are summarized in Table \ref{tab:table1}.

\begin{figure*}[tbp]
  \centering
  \includegraphics[width=0.8\textwidth,trim=1.5 1.5 1.5 1.5,clip]{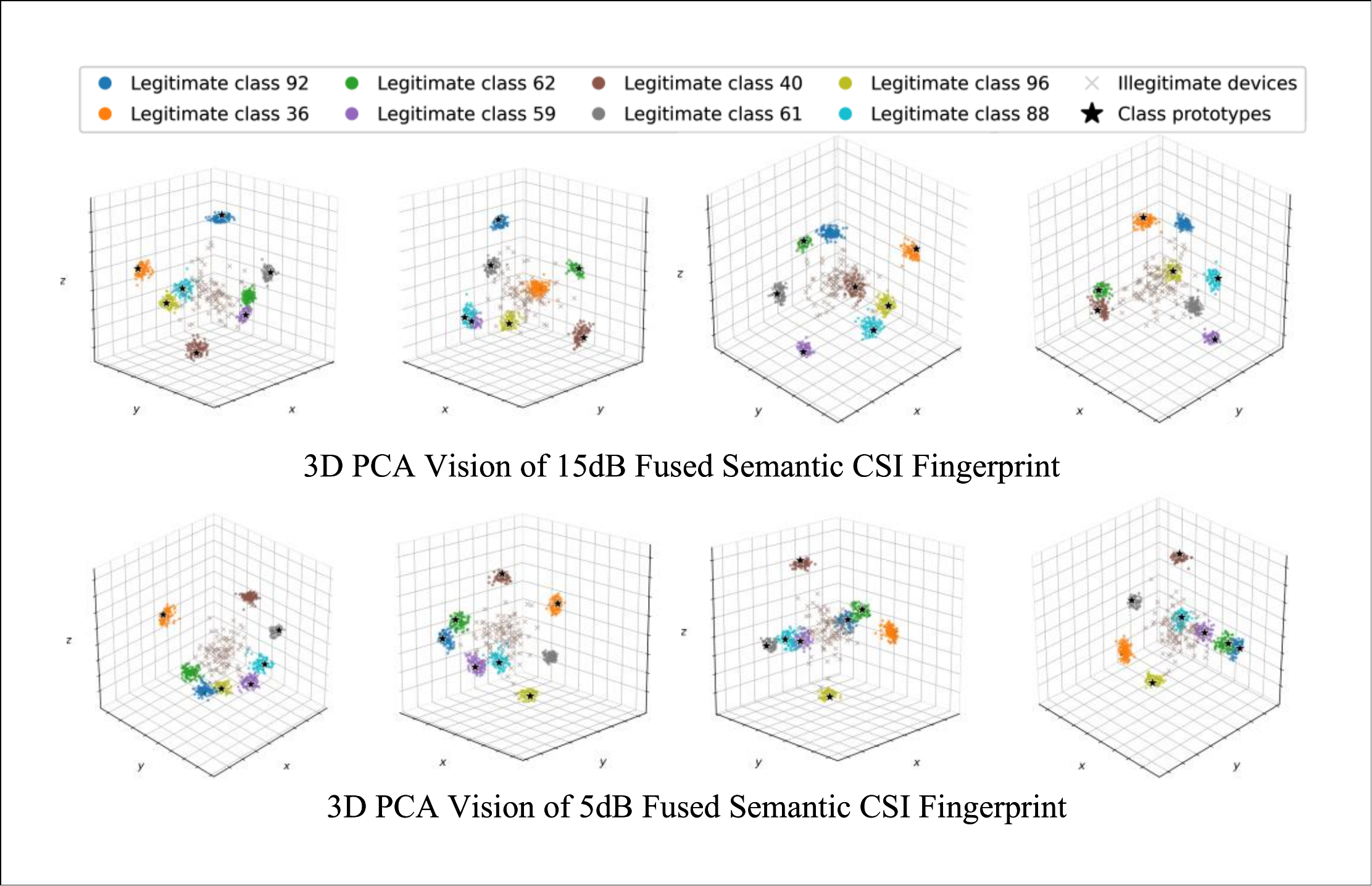}
  \caption{The 3D PCA visualization of the fused semantic CSI fingerprints under different SNRs. The upper row shows the fused feature distribution at 15 dB, and the lower row shows the fused feature distribution at 5 dB. The colored dots denote legitimate samples from different enrolled device classes, the cross markers denote illegal samples, and the black stars denote class prototypes. The four subfigures in each row present the same PCA feature space from different viewing angles, where $x$, $y$, and $z$ represent the first three principal components.}
  \label{15dB_pca}
\end{figure*}

The resulting dataset is used to build both the closed-set classification task and the open-set authentication task. In the closed-set setting, the 100 legitimate devices are treated as 100 identity classes. In the open-set setting, the legitimate-device samples are first authenticated to reject impersonating attackers and are then forwarded to the legitimate-device classifier. To align with this protocol, the CSI sequences of legitimate devices collected over the 500 time slots are randomly divided into training and test subsets with a ratio of 4:1, whereas attacker samples are used only during evaluation to assess spoofing resistance. For the distributed semantic authentication component, the semantic encoder is first pre-trained with the reconstruction-oriented objective in \eqref{eq:semantic_pretrain_loss_final} and then transferred to initialize the collaborative training stage of fusion and authentication.

\subsection{Comparative Schemes}
We consider the comparative baselines as follows:


\begin{itemize}
  \item \textbf{Single Node based Semantic PLA \cite{jing2023multi}:}
    Only central node Bob's local semantic fingerprint is used for authentication and classification, without any distributed cooperation.
  \item
    \textbf{Distributed Hard Voting PLA \cite{zhang2023distributed}:}
    Each distributed node makes an individual class decision, and the final result is determined by majority voting at Bob.
  \item
    \textbf{Distributed Traditional Transmission PLA (D2T-PLA) \cite{mahmood2017channel}:}
    Distributed nodes transmit raw CSI to Bob through a conventional communication pipeline rather than task-oriented semantic transmission.
  \item
    \textbf{Mean Feature Fusion Semantic PLA:}
    The features from different distributed nodes and central receiver Bob are first extracted and then directly averaged before final authentication.
  \item
    \textbf{Feature Concat MLP Semantic PLA \cite{yan2025self}:}
    Features from all nodes are concatenated and then fed into an MLP for joint decision making.

\end{itemize}

\begin{figure}[tbp]
  \centering
  \includegraphics[width=0.40\textwidth,trim=0.2 0.2 0.2 0.2,clip]{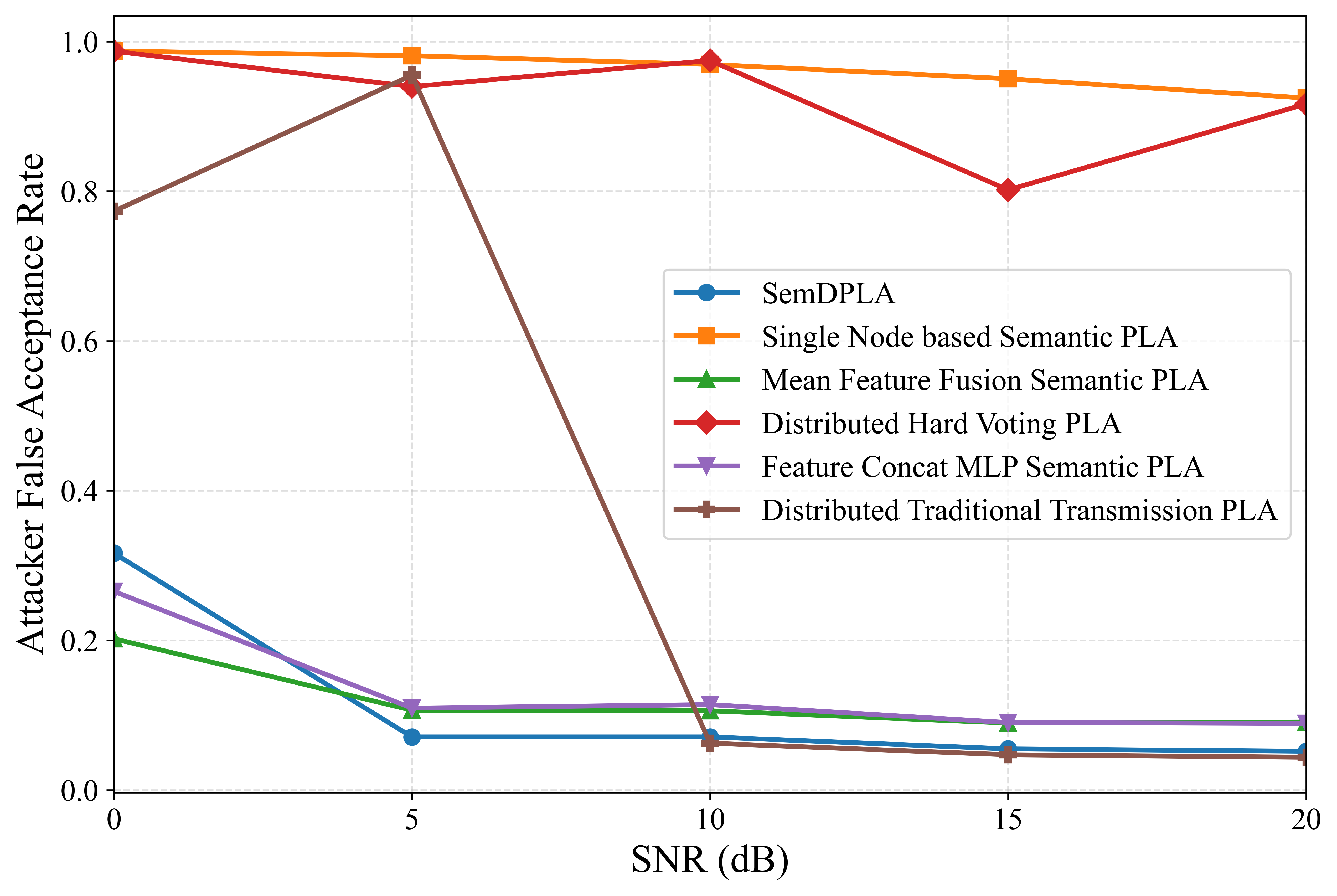}
  \caption{Comparison of attacker false acceptance rates under different SNRs when the false rejection rate is constrained to be no higher than 1\%.}
  \label{fig1_attacker_filtering_capability}
\end{figure}

\subsection{Simulation Results of Fingerprint Prediction}


Fig.~\ref{15dB_pca} shows a visualization of the 3D PCA of the semantic fusion fingerprints at 15 dB and 5 dB. The colored points represent valid samples from different categories, black stars denote the centers of the categories, and cross marks indicate invalid samples. As can be seen from the figures at different angles, under different signal-to-noise ratios, valid samples cluster closely around their respective category centers, and the different categories remain well separated in the projection space. Furthermore, invalid samples are also located far from the valid clusters. This shows that the fused features generated by our approach have good discriminability.

Fig. \ref{fig1_attacker_filtering_capability} compares the false acceptance rate (FAR) of different schemes under fixed constraints. The proposed SemDPLA maintains a low FAR across different SNR levels, indicating that SemDPLA exhibits great robustness in authentication.
Under high SNR conditions, the performance of mean feature fusion and feature fusion MLP consistently lags behind that of the proposed scheme. This is because fixed feature averaging or direct feature concatenation fails to capture the mutual information differences in distributed semantic fingerprints, thereby preventing distributed nodes, which provide more valuable feature supplementation to the central node,  from playing a greater role in the fusion process.
Furthermore, single-node semantic PLA and distributed hard-voting PLA exhibit high FARs across the entire SNR range, indicating that single-node fingerprinting and voting decisions are insufficient for reliable authentication in dense device scenarios. Distributed conventional PLA performs well at high SNR, but its FAR significantly increases at 0 dB and 5 dB, demonstrating its sensitivity to degraded links. In contrast, SemDPLA offers more stable performance at low SNR, demonstrating the robustness of semantic transmission and adaptive fusion under non-ideal channel conditions.



\begin{figure}[t]
  \centering
  \includegraphics[width=0.4\textwidth,trim=0.2 0.2 0.2 0.2,clip]{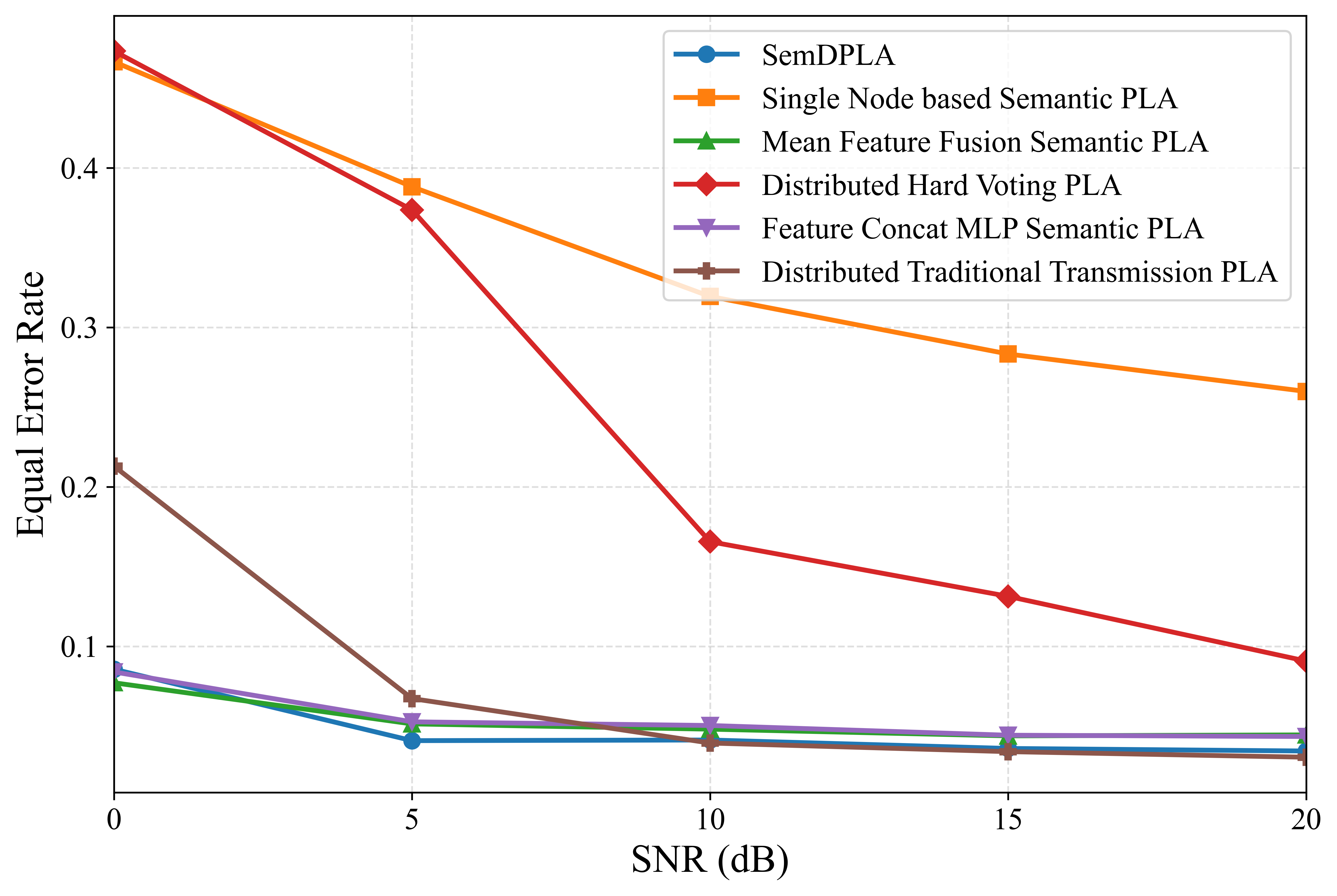}
  \caption{Comparison of equal error rates under different SNRs. A lower EER indicates better balanced authentication performance between illegal-device acceptance and legitimate-device rejection.}
  \label{fig_eer_vs_snr}
\end{figure}

\begin{figure}[t]
  \centering
  \includegraphics[width=0.4\textwidth,trim=0.2 0.2 0.2 0.2,clip]{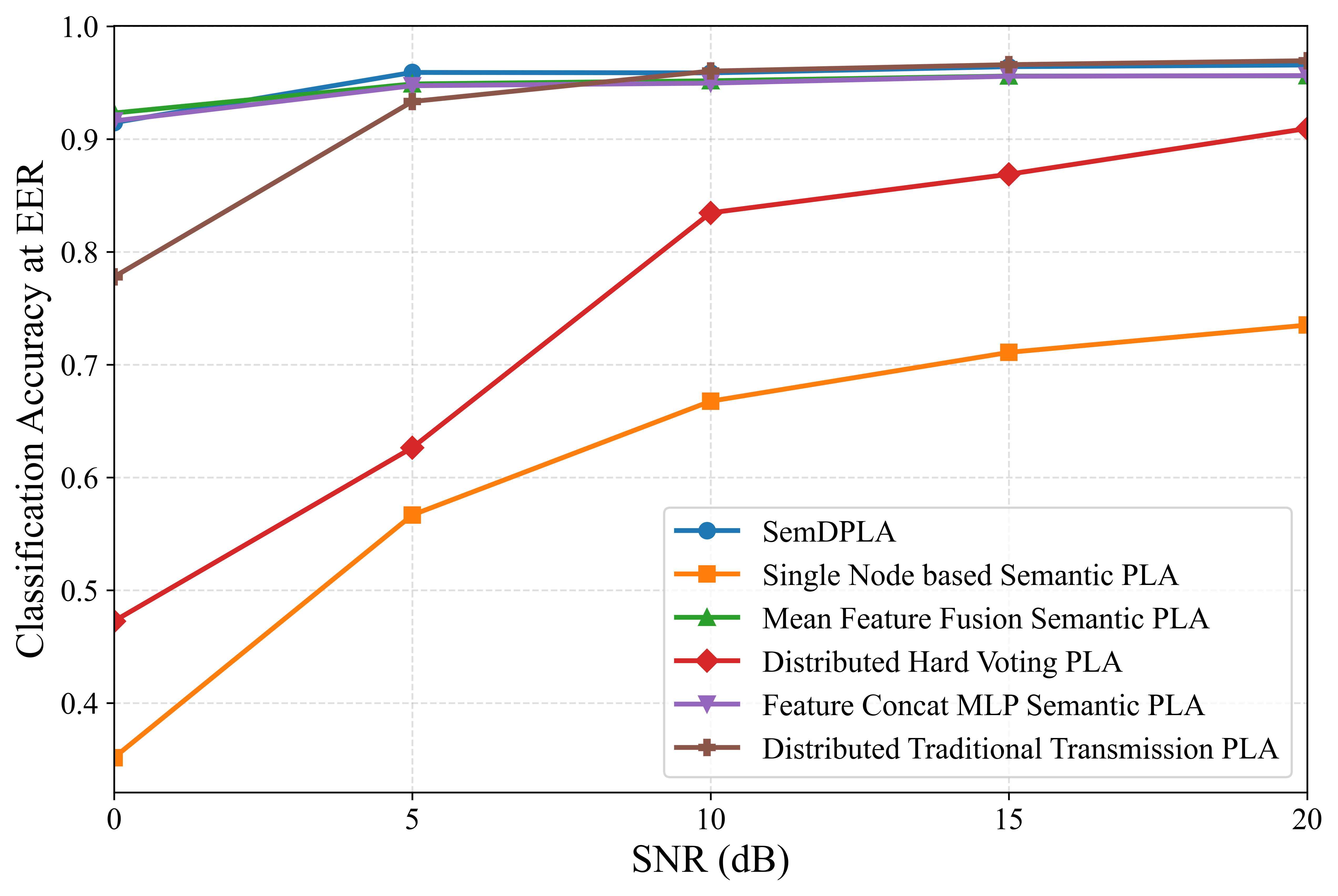}
  \caption{Comparison of classification accuracy at the EER operating point under different SNRs. The accuracy is measured on legitimate-device classification under the threshold where FAR and FRR are approximately equal. A higher value indicates better closed-set identification capability under balanced authentication.}
  \label{fig_auth_accuracy_at_eer}
\end{figure}

Fig. ~\ref{fig_eer_vs_snr} compares the equal error rate (EER) of the proposed SemDPLA with that of various baseline schemes under different SNR conditions. The EER corresponds to an operating point where the FAR and FRR are approximately equal. SemDPLA maintains a low EER across the entire SNR range, decreasing from approximately $8.6\%$ at 0 dB to approximately $3.4\%$ at 20 dB, indicating a stable authentication threshold across varying channel qualities.
Compared to average feature fusion and feature-consistency MLP, SemDPLA achieves a lower EER from 5 dB onwards in most cases.
The single-node semantic PLA maintains a high EER across all SNR levels, remaining at around $26\%$ even at 20 dB, suggesting that reliance on a single fingerprint is insufficient for reliable authentication. The distributed hard-voting PLA improves as SNR increases, but its EER remains close to $47\%$ and $37\%$ at 0 dB and 5 dB respectively, indicating that voting decisions are susceptible to local errors under low SNR conditions. The distributed traditional transmission PLA performs well at 10 dB and above, primarily because high-dimensional feature transmission suffers less distortion when channel quality is good. However, its EER increases significantly at 0 dB and 5 dB, demonstrating greater sensitivity to transmission link degradation. In contrast, SemDPLA provides more stable authentication performance at low SNR through semantic compression and adaptive fusion.


\begin{figure}[t]
  \centering
  \includegraphics[width=0.4\textwidth,trim=0.2 0.2 0.2 0.2,clip]{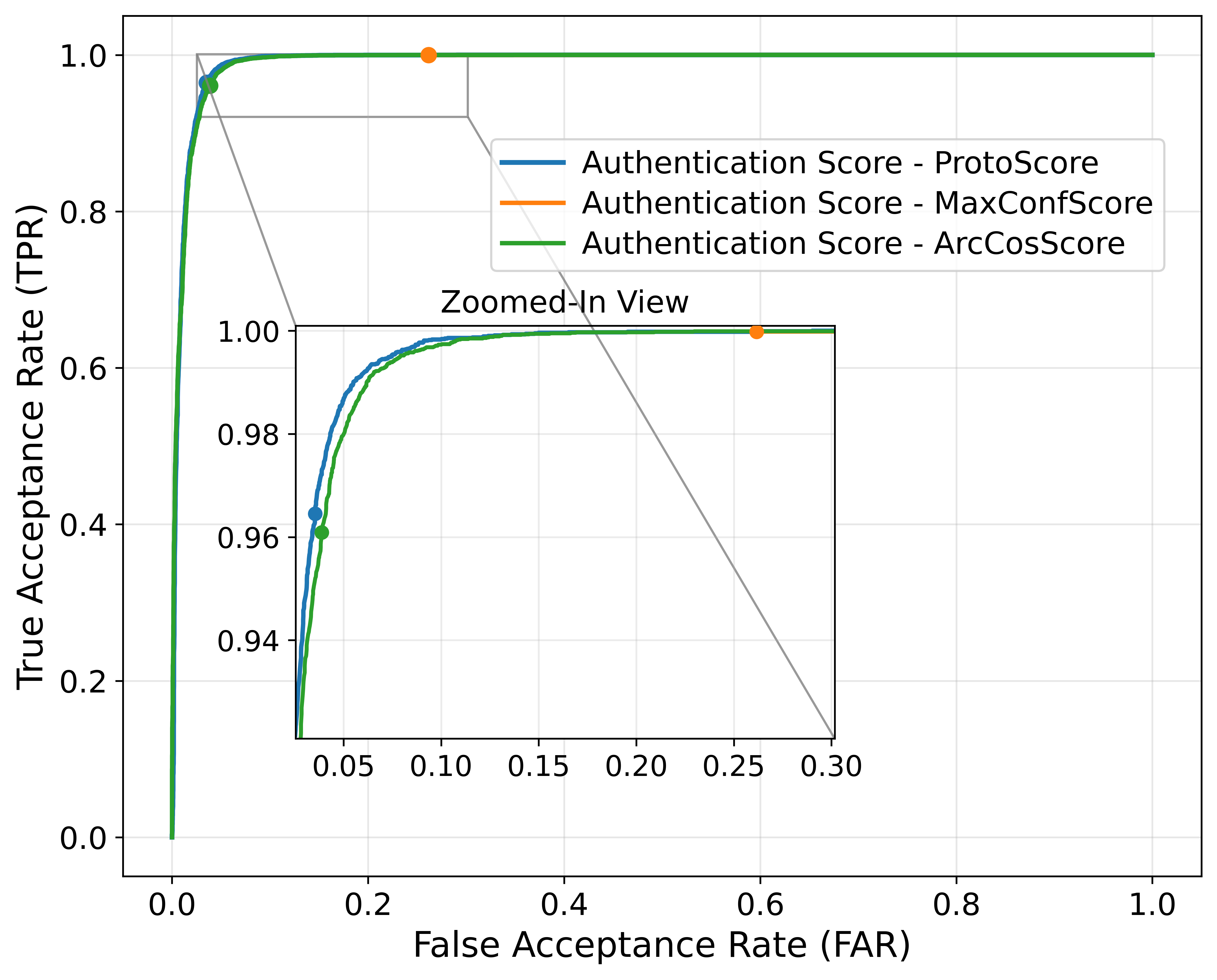}
  \caption{ROC curves of ProtoScore, MaxConfScore, and ArcCosScore at 15 dB.}
  \label{fig3_three_scores_roc_three_methods_with_inset}
\end{figure}

\begin{figure}[t]
  \centering
  \includegraphics[width=0.4\textwidth,trim=0.2 0.2 0.2 0.2,clip]{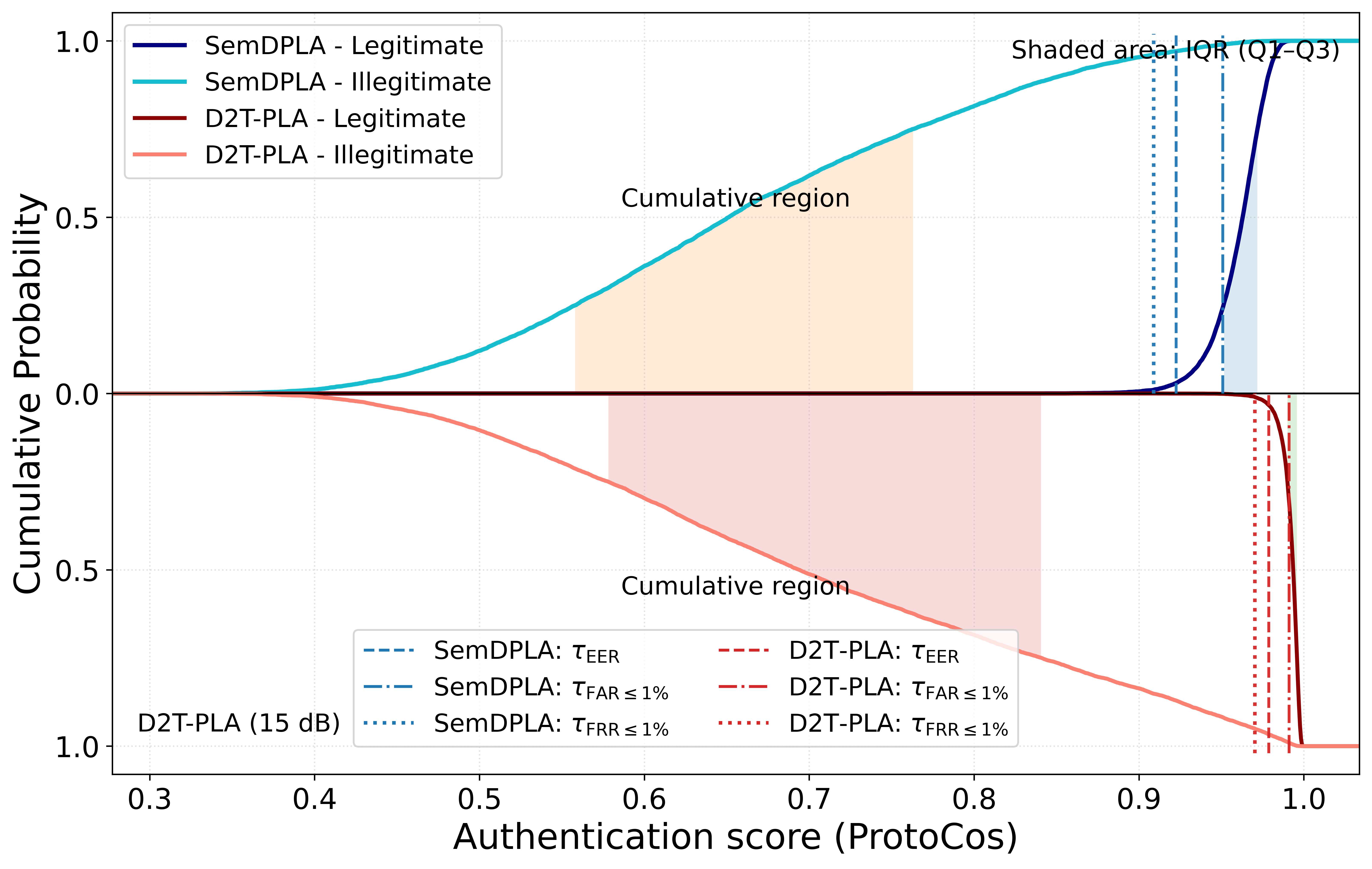}
  \caption{CDFs of ProtoCos authentication scores for SemDPLA and D2T-PLA at SNR=5 dB. The upper and lower parts correspond to SemDPLA and D2T-PLA, respectively. The shaded regions indicate the score intervals between [0.25, 0.75] of legitimate and illegitimate samples.}
  \label{mirror_cdf_15dB}
\end{figure}

\begin{figure}[t]
  \centering
  \includegraphics[width=0.4\textwidth,trim=0.2 0.2 0.2 0.2,clip]{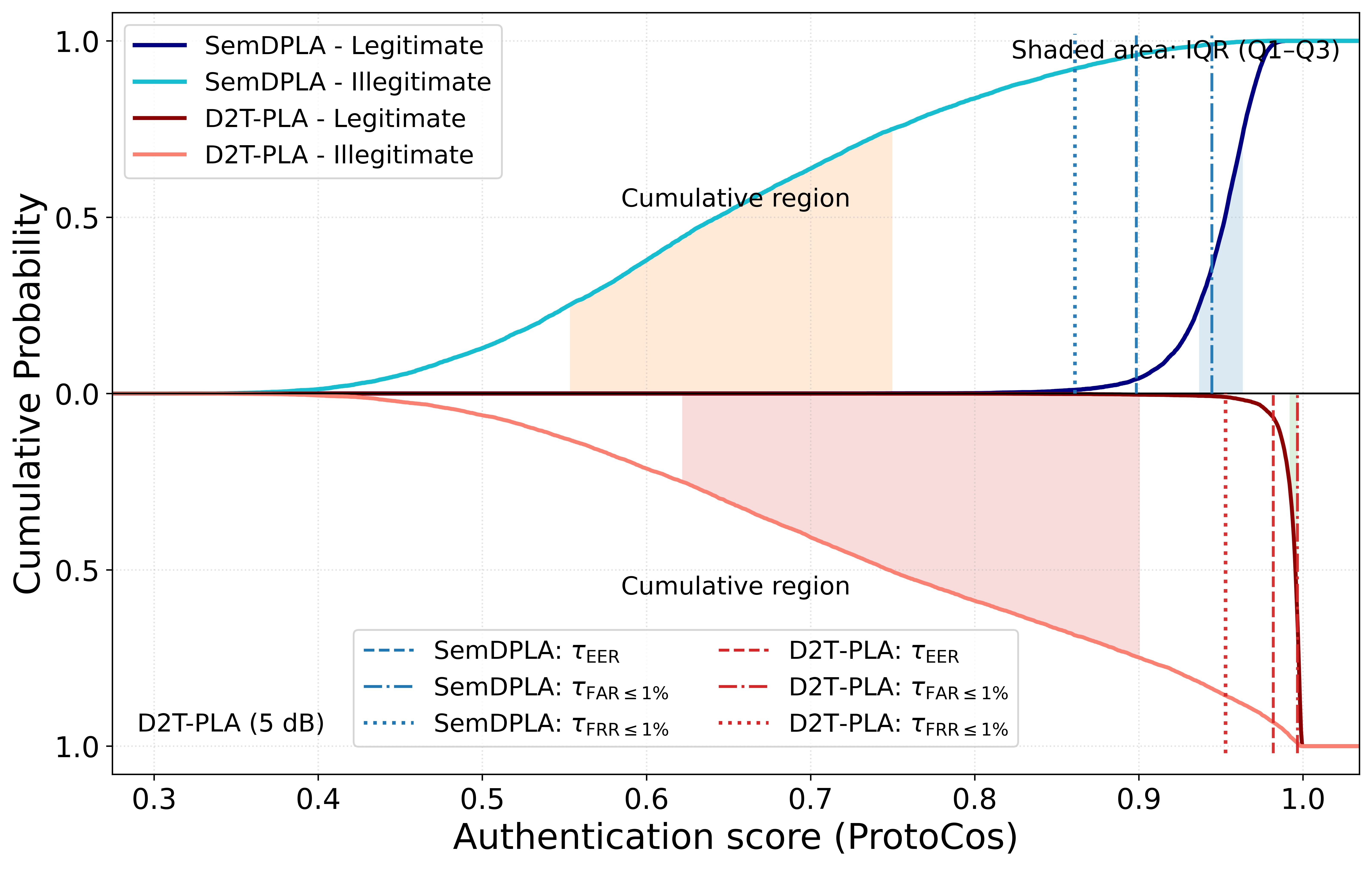}
  \caption{CDFs of ProtoCos authentication scores for SemDPLA and D2T-PLA at SNR=15 dB. The upper and lower parts correspond to SemDPLA and D2T-PLA, respectively. The shaded regions indicate the score intervals between [0.25, 0.75] of legitimate and illegitimate samples}
  \label{mirror_cdf_5dB}
\end{figure}

Fig. ~\ref{fig_auth_accuracy_at_eer} shows the legitimate device classification accuracy of different schemes at the EER point. As the SNR increases, the classification accuracy of all schemes improves. Together with Fig. ~\ref{fig_eer_vs_snr}, it can be observed that SemDPLA achieves an accuracy of $91.4\%$ at 0 dB and maintains an accuracy of over $95.8\%$ between 5 dB and 20 dB. Our proposed scheme not only maintains a low balanced authentication error but also maintains a high closed-set recognition accuracy at the corresponding EER thresholds.
In contrast, the single-node semantic PLA is more sensitive to SNR, increasing from approximately $35.2\%$ at 0 dB to approximately $73.5\%$ at 20 dB, yet it remains significantly lower than that of the distributed fusion scheme. The distributed hard-voting scheme demonstrates superior overall classification performance compared to the single-node semantic PLA scheme; however, it performs poorly under low SNR conditions, indicating that voting remains more susceptible to the influence of local erroneous decisions, leading to classification inaccuracies.
The performance of mean feature fusion and feature consistency MLP approaches that of SemDPLA.
The distributed traditional transmission PLA achieves accuracy comparable to or slightly higher than that of SemDPLA when SNR exceeds 10 dB. This is because high-dimensional feature transmission suffers less distortion under favourable channel conditions, allowing for the preservation of more classification details. However, its accuracy drops to approximately $77.7\%$ at 0 dB, demonstrating greater sensitivity to transmission-link degradation. Overall, SemDPLA offers more stable classification performance at low SNR whilst maintaining extremely high accuracy at high SNR.

Fig. ~\ref{fig3_three_scores_roc_three_methods_with_inset} compares the receiver operating characteristic (ROC) curves of ProtoScore, MaxConfScore and ArcCosScore at 15 dB. The ROC curve reflects the true acceptance rate of legitimate devices at different false acceptance rates for unauthorised devices; the closer the curve is to the top-left corner, the stronger the discriminatory power of the authentication score.
The ROC curves demonstrate that ProtoScore and ArcCosScore significantly outperform MaxConfScore, indicating that scores based on cosine similarity are better suited to open-set authentication tasks, whereas closed-set softmax confidence scores struggle to reliably reject unauthorised devices. In the magnified section showing low false acceptance rates, ProtoScore performs slightly better overall than ArcCosScore. This indicates that, for the same false acceptance rate of unauthorised devices, ProtoScore achieves a higher acceptance rate for authorised devices. Consequently, this paper selects ProtoScore as the authentication score for SemDPLA, to be used for subsequent threshold-based authentication.

Figs.~\ref{mirror_cdf_15dB} and~\ref{mirror_cdf_5dB} compare the mirrored CDFs of ProtoCos authentication scores for SemDPLA and D2T-PLA at 5 dB and 15 dB. At 5 dB, the distributions of legitimate and illegitimate devices in SemDPLA are clearly separated. The central $50\%$ interval of the legitimate scores is concentrated in the high-score region, while the central interval of the illegitimate scores lies in the lower range. This indicates that the proposed method can still maintain a high degree of discrimination between legitimate and illegitimate devices at low SNR. In contrast, D2T-PLA exhibits a more widely distributed score distribution. The illegal scores extend into the high-score region,  resulting in partial confusion between legal and illegal devices. This leads to poor authentication quality during transmission at low SNR.
At 15 dB, SemDPLA compresses the legal scores towards 1.0 while maintaining a clear separation of the illegal scores. This provides a stable score gap for threshold selection. However, for D2T-PLA, although legitimate scores are also highly concentrated around 1.0, the decision threshold is pushed into a very narrow high-score range, which indicates that the scheme becomes sensitive to small threshold shifts or score perturbations. During testing, even minor threshold deviations can lead to a significant decline in authentication performance.
Overall, SemDPLA provides a more stable score distribution under both SNR settings; at low SNR, it reduces the overlap between legitimate and illegitimate devices. At high SNR, it avoids an excessively narrow valid threshold region.

\begin{figure}[t]
  \centering
  \includegraphics[width=0.4\textwidth,trim=0.2 0.2 0.2 0.2,clip]{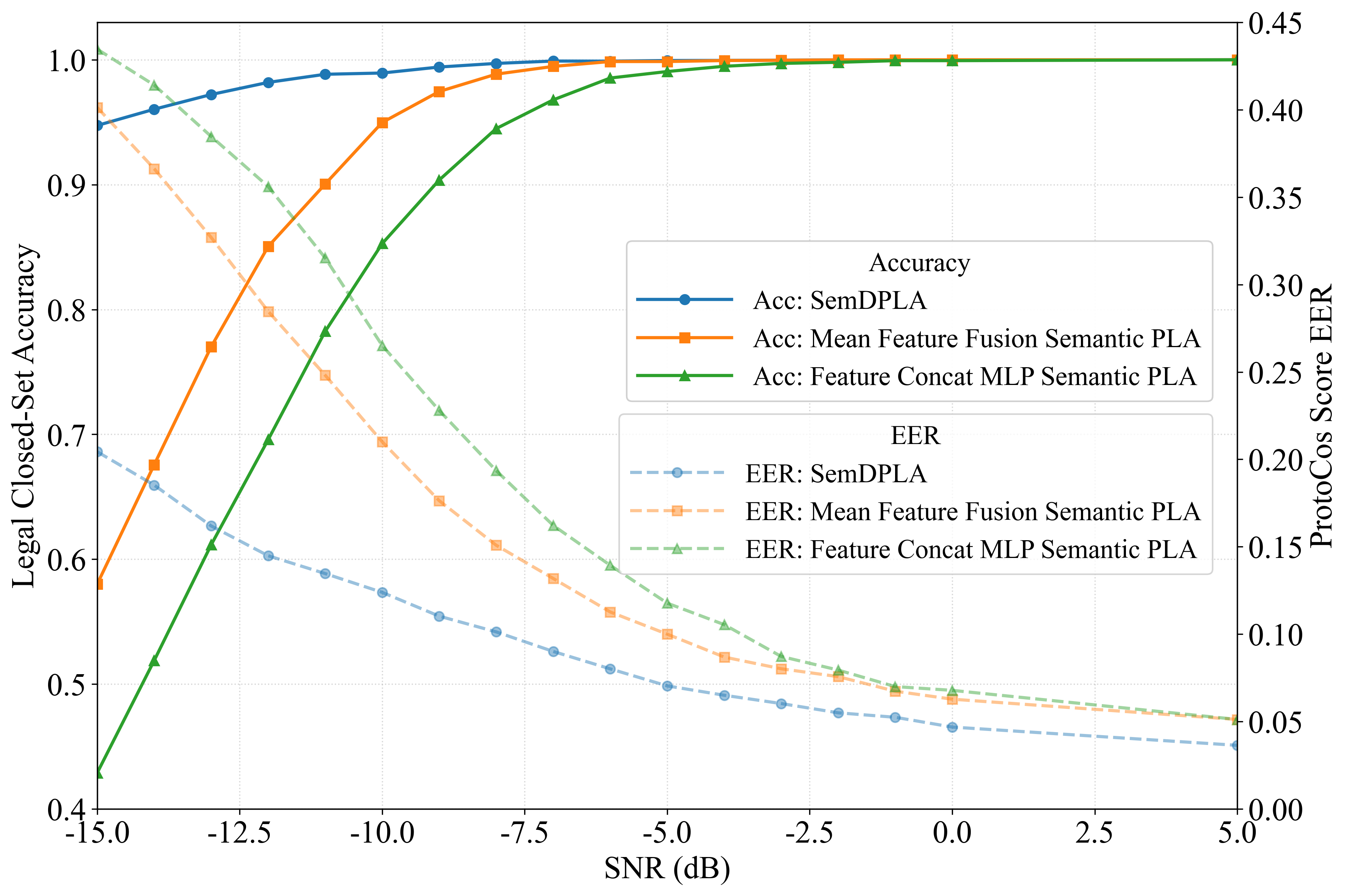}
  \caption{Accuracy and EER performance under abnormal-node conditions. The environment SNR=15 dB, while the horizontal axis denotes the SNR of the abnormal node.}
  \label{dual_axis_optimized_internal}
\end{figure}

\begin{figure}[t]
  \centering
  \includegraphics[width=0.4\textwidth,trim=0.2 0.2 0.2 0.2,clip]{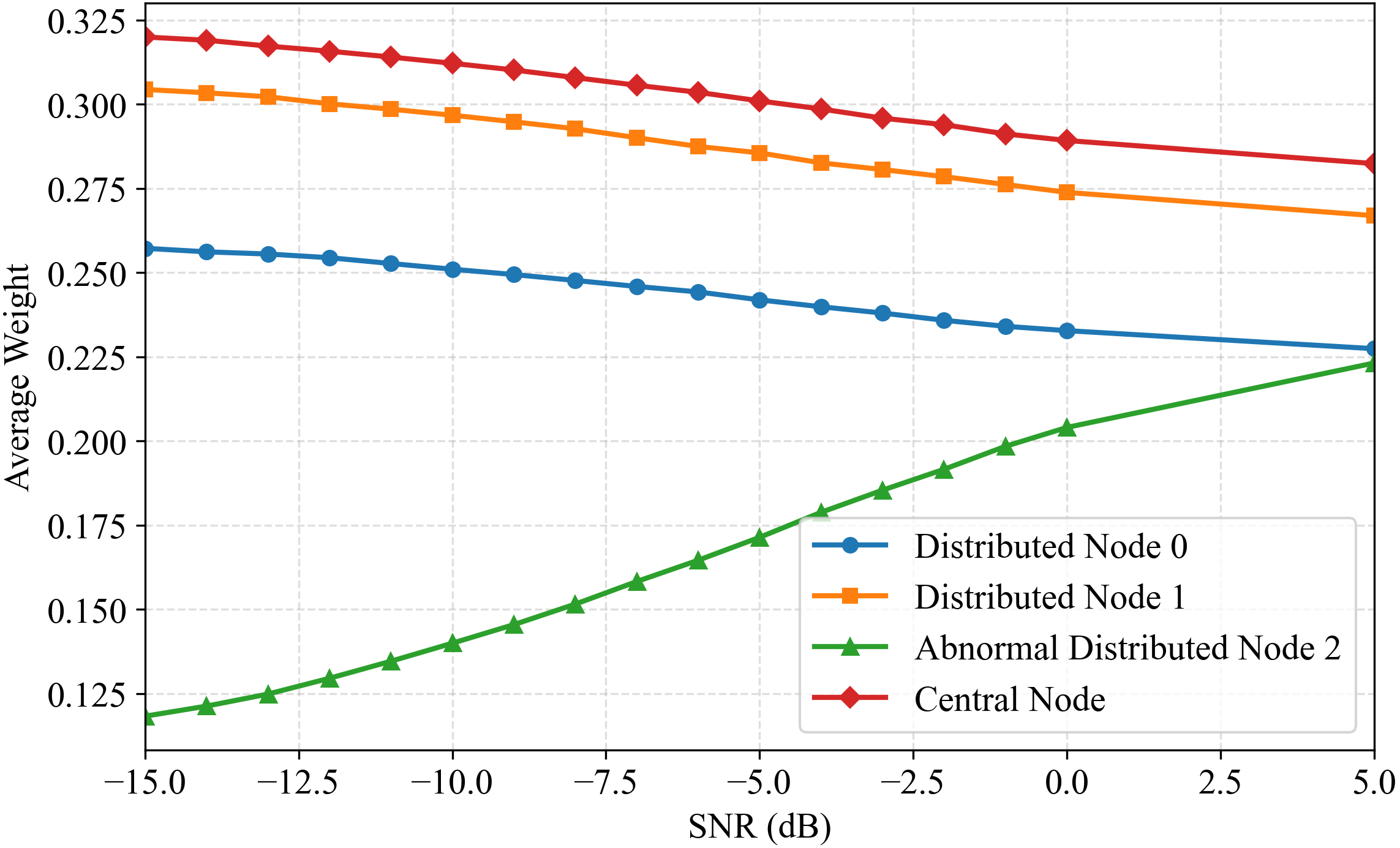}
  \caption{Average fusion weights under abnormal-node conditions. The horizontal axis denotes the SNR of the abnormal node, while the other distributed nodes and the central observation are at environment SNR=15 dB. The four curves represent the average weights assigned to Node 0, Node 1, the corrupted Node 2, and the central semantic CSI fingerprints.}
  \label{avg_weight_vs_snr}
\end{figure}

Fig.~\ref{dual_axis_optimized_internal} illustrates how the authentication performance of different schemes is affected when an abnormal node is present in the receiver. As can be seen from the figure, as the SNR of the attacked node increases, the classification accuracy of the proposed SemDPLA scheme and the two comparison schemes improves, and the equal error rate decreases accordingly. However, when the semantic fingerprint information from abnormal nodes is unreliable, the authentication and classification performances of Mean Feature Fusion Semantic PLA and Feature Concat MLP Semantic PLA are far inferior to that of the proposed scheme. This indicates that semantic fingerprints based on adaptive fusion of node channel quality are more resistant to attacks from abnormal nodes than schemes based on weighted average fusion or feature concatenation.



Fig.~\ref{avg_weight_vs_snr} further illustrates the adaptive fusion process of the proposed SemDPLA scheme in the event of anomalous nodes. As can be seen from the figure, when the SNR of the attacked node is very low, the model reduces the fusion weight assigned to that distributed node, while the fusion weights of the remaining nodes are correspondingly increased. Furthermore, as the SNR of the attacked node improves and the quality of its fingerprint increases, the model gradually assigns the attacked node a higher fusion weight to enhance its influence on the fused fingerprint. In addition, the figure shows that the weight of the central node remains higher than that of the other distributed nodes, and the weights of the distributed nodes are not distributed uniformly. This indicates that the central node possesses higher-quality features and the fusion is centred around the central node. Moreover, the influence of different nodes’ fingerprints on the fused fingerprint varies. Consequently, adaptive weighting enhances the quality of the fusion, as illustrated in Fig.~\ref{dual_axis_optimized_internal}.

\section{Conclusion}
\label{sec:conclusion}

We have proposed SemDPLA to enable multi-device authentication in high-density 6G IoT scenarios. In this scheme, we utilize DPLA to identify devices in high-density environments. Furthermore, semantic communication is introduced to ensure communication quality while reducing communication resource consumption. For fingerprint design, we enhance the system’s resistance to attacks by anomalous nodes through the adaptive fusion of distributed semantic fingerprints. The classification mechanism of ArcFace further widens the classification boundaries for devices. The proposed framework demonstrates significant potential for deployment in 6G high-density IoT scenarios.

\bibliography{ref.bib}
\bibliographystyle{IEEEtran}

\vfill

\end{document}